\documentclass[%
 reprint,
 superscriptaddress,
nofootinbib,
 amsmath,amssymb,
 aps,
 pre,
 floatfix,
 nobalancelastpage
]{revtex4-2}
\PassOptionsToPackage{dvipsnames}{xcolor}
\usepackage{graphicx}% Include figure files
\usepackage{dcolumn}% Align table columns on decimal point
\usepackage{bm}% bold math
\usepackage{braket}
\usepackage{enumitem}
\usepackage{physics}
\usepackage{extarrows}
\usepackage{tikz}
\usepackage[caption=false]{subfig}
\newcommand{\phantomsubfloat}[1]{
    {% apply caption setup only temporarily
        \captionsetup[subfigure]{labelformat=empty}
        \subfloat[][]{#1}
    }%
}

\usepackage{hyperref}
\usepackage{derivative}
\usepackage{comment}
\usepackage{amsthm}
\usepackage[capitalise]{cleveref}
\hypersetup{colorlinks={true},allcolors={blue}}

\usepackage{appendix}

\newcommand{\beginsupplement}{
  \setcounter{table}{0}  
  \renewcommand{\thetable}{A\arabic{table}} 
  \setcounter{figure}{0} 
  \renewcommand{\thefigure}{A\arabic{figure}}
}

\binoppenalty=\maxdimen
\relpenalty=\maxdimen

\definecolor{royalblue}{rgb}{0.2, 0.14, 0.4}

\newcommand*\diff{\mathop{}\!\mathrm{d}}
\newcommand{\ddiff}[1]{\mathop{}\!\frac{\mathrm{d}#1}{\mathrm{d}t}}
\newcommand{\iu}{\mathrm{i}\mkern1mu}

\usepackage{booktabs}
\usepackage{nicematrix}
\definecolor{ylw}{RGB}{250,234,203}
\newcommand{\Input}[1]{\textcolor{RoyalBlue!85!black}{#1}}
\newcommand{\Output}[1]{\textcolor{BrickRed}{#1}}

\begin{document}

\title{Effective One-Dimensional Reduction of Multicompartment Complex Systems Dynamics}
%%% AUTHORS and AFFILIATIONS
\author{Giorgio Vittorio Visco$^\mathsection$}
\email{giorgiovittorio.visco@phd.unipd.it}
\affiliation{Department of Physics and Astronomy ``Galileo Galilei'', University of Padua, Via F. Marzolo 8, 35131, Padova, Italy}

\author{Johannes Nauta$^\mathsection$}
\affiliation{Dept. of Physics and Astronomy ``Galileo Galilei'', University of Padua, Via F. Marzolo 8, 35131, Padova, Italy}

\author{Tomas~Scagliarini}
\affiliation{Dept. of Physics and Astronomy ``Galileo Galilei'', University of Padua, Via F. Marzolo 8, 35131, Padova, Italy}

\author{Oriol Artime$^{\ddagger}$}
\affiliation{Departament de Física de la Matèria Condensada, Universitat de Barcelona, 08028 Barcelona, Spain}
\affiliation{University of Barcelona Institute of Complex Systems (UBICS), Universitat de Barcelona, 08028 Barcelona, Spain}
\affiliation{Universitat de les Illes Balears, 07122 Palma, Spain}

\author{Manlio De Domenico$^{\dagger,\ddagger,}$}
\affiliation{Dept. of Physics and Astronomy ``Galileo Galilei'', University of Padua, Via F. Marzolo 8, 35131, Padova, Italy}
\affiliation{Padua Center for Network Medicine, University of Padua, Via F. Marzolo 8, 35131 Padova, Italy}
\affiliation{Istituto Nazionale di Fisica Nucleare, Sezione Padova, Italy}
\email{manlio.dedomenico@unipd.it}

\date{\today}% It is always \today, today,
             % but any date may be explicitly specified

\def\thefootnote{$\mathsection$}%
\footnotetext{These authors contributed equally to this work.}%
\def\thefootnote{$\ddagger$}%
\footnotetext{These authors jointly supervised this work.}%
\def\thefootnote{\arabic{footnote}}

%%% ABSTRACT
\begin{abstract}
A broad class of systems, including ecological, epidemiological, and sociological ones, are characterized by populations of individuals assigned to specific categories, e.g., a chemical species, an opinion, or an epidemic state, that are modeled as compartments. Because of interactions and intrinsic dynamics, the system units are allowed to change category, leading to concentrations varying over time with complex behavior, typical of reaction-diffusion systems. While compartmental modeling provides a powerful framework for studying the dynamics of such populations and describe the spatiotemporal evolution of a system, it mostly relies on deterministic mean-field descriptions to deal with systems with many degrees of freedom.
Here, we propose a method to alleviate some of the limitations of compartmental models by capitalizing on tools originating from quantum physics to systematically reduce multidimensional systems to an effective one-dimensional representation. Using this reduced system, we are able not only to investigate the mean-field dynamics and their critical behavior, but we can additionally study stochastic representations that capture fundamental features of the system. We demonstrate the validity of our formalism by studying the critical behavior of models widely adopted to study epidemic, ecological and economic systems.
\end{abstract}

%%% TITLE
\maketitle

%%% MANUSCRIPT
\section{Introduction}\label{sec:introduction}
Complex systems consist of a multitude of interconnected and interacting entities that form extensive networks, with typical examples ranging from chemical to ecological and social ones~\cite{freeman2004development,montoya2006ecological,kossinets2006empirical,fath2007ecological,castellano2009statistical,ings2009ecological,bascompte2010structure,de2013anatomy,gallotti2020assessing}. To analyze their complex behavior, often characterized by collective dynamics, phase transitions, and other emergent phenomena, compartmental models are usually adopted. 
In this framework, units of the system are categorized or stratified into distinct species, states, or groups, referred to as compartments. Then, the concentration of units in each compartment is studied over time, taking into account the change of species due to either unit-unit interactions or spontaneous transitions into other compartments. 

Compartmentalization facilitates the study of dynamic behaviors and phase transitions in some order parameters as a function of control parameters that directly depend on modeling assumptions and observables. For instance, in the case of an epidemic outbreak, it is possible to study how the steady-state density of infected individuals depends on the transmission rate of a pathogen by means of social contacts and the recovery rate from the corresponding disease~\cite{belik2011natural,gomez2018critical,castioni2021critical} (see also Ref.~\cite{pastor2015epidemic} for an extensive review). One of the most successful frameworks to study the behavior of such epidemic systems, as well as those consisting of multiple interconnected compartments, is based on the study of the next-generation matrix, an indispensable tool to analyze the conditions required to favor the emergence of active states where epidemics dynamics is sustained and infection has the potential to scale at system level~\cite{diekmann1990definition,diekmann2010construction}.

%% FIGURE
\begin{figure*}[t]
    \centering
    \resizebox{\linewidth}{!}{%
    \begin{tikzpicture}[every node/.style={inner sep=.5pt}]
        \node (a) {%
            \includegraphics[height=.25\textheight]{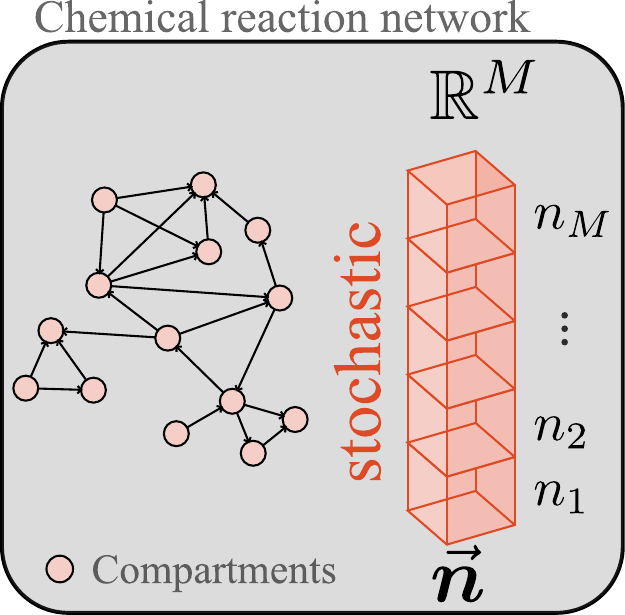}
        };
        \node[anchor=north west,yshift=-4.5ex,xshift=1em] (alabel) at (a.north west) {\LARGE (a)};
        \node[anchor=west] (b) at (a.east) {%
            \includegraphics[height=.25\textheight]{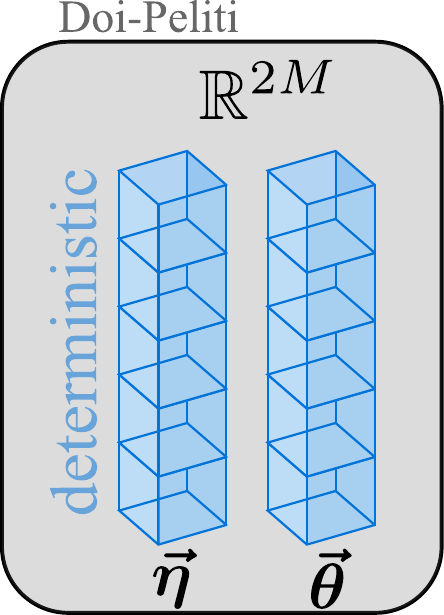}
        };
        \node[anchor=west,xshift=1em] (blabel) at (alabel -| b.north west) {\LARGE (b)};
        \node[anchor=west] (c) at (b.east) {%
            \includegraphics[height=.25\textheight]{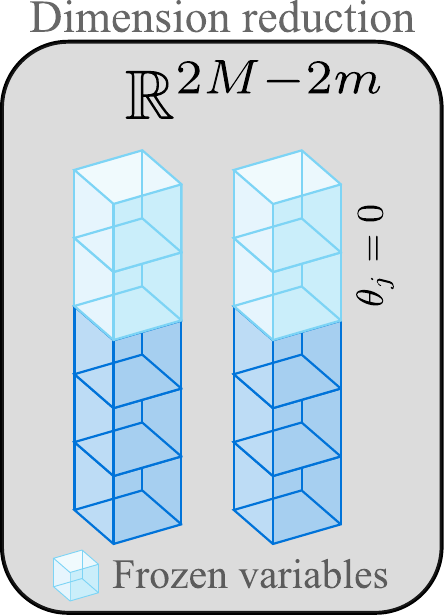}
        };
        \node[anchor=west,xshift=1em] (clabel) at (blabel -| c.north west) {\LARGE (c)};
        \node[anchor=west] (d) at (c.east) {%
            \includegraphics[height=.25\textheight]{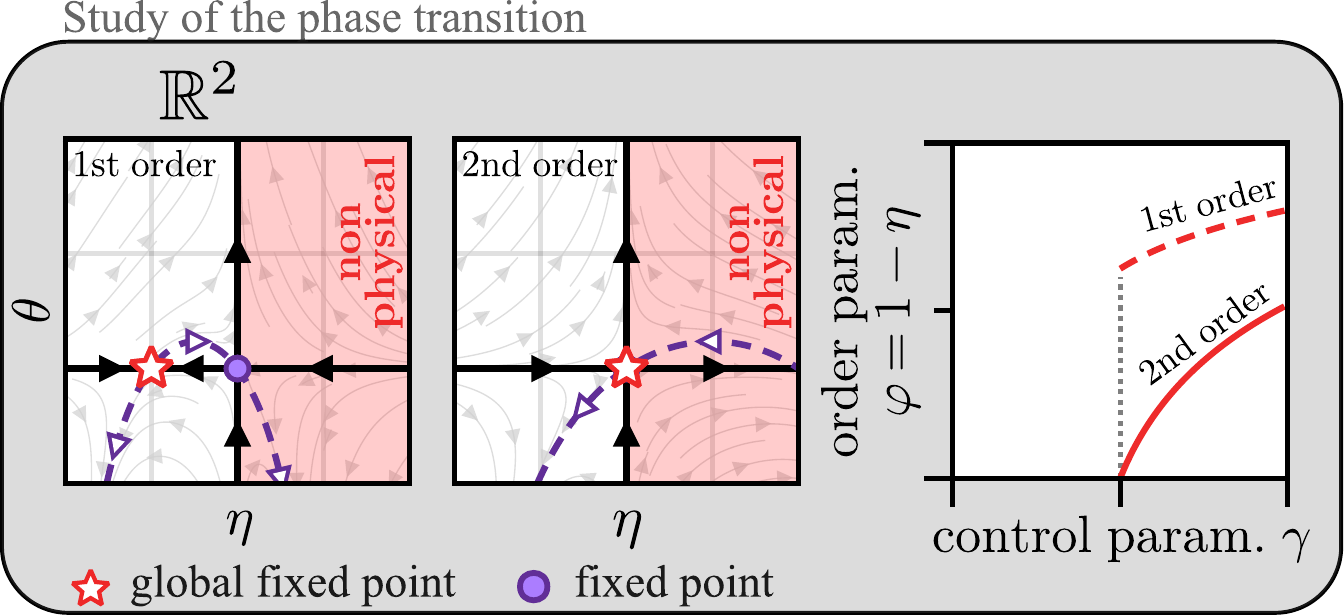}
        };
        \node[anchor=west,xshift=1em] (dlabel) at (clabel -| d.north west) {\LARGE (d)};
        \phantomsubfloat{\label{fig:intro:a}}
        \phantomsubfloat{\label{fig:intro:b}}
        \phantomsubfloat{\label{fig:intro:c}}
        \phantomsubfloat{\label{fig:intro:d}}        
    \end{tikzpicture}
    }
    % \end{subcaptiongroup}
    \caption{%
    \textbf{Dimension reduction of the mathematical description of chemical reaction networks and characterization of their phase transitions.} 
    We transform chemical reaction networks consisting of $M$ compartments~(a), for which its stochastic dynamics are described by the master equation (\cref{eq:ME}), into a deterministic Hamiltonian system of dimension $2M$ (Eq.~\eqref{eq:h2sist}) using the Doi-Peliti formalism (b). 
    Then, we introduce $m$ constraints by ``freezing'' some compartments (\cref{eq:integraliprimi}) in order to reduce the dimensions of the system (c).
    Using this reduced system, we can study phase transitions that the original system exhibits by varying a control parameter $\gamma$, which determines the position of the fixed points in the phase portrait (d). 
    Here, the phase portraits (d, left and middle) are depicted at exactly the critical point $\gamma_c$ (d, right, dashed line), and the obtained reduced dynamics are captured by the occupation number $\eta$ and the noise variable $\theta$ (see~\cref{eq:colehopf}).
    The shaded red areas in (d) are the nonphysical regions, as they correspond to regions that cannot be attained --- for example, occupation numbers larger than the total population.
    The (global) fixed points are on the intersections of the trivial zero-energy lines (black) and the nontrivial zero-energy lines (dashed purple).
    When the phase transition is a first-order transition, the order parameter ``jumps'' to a nonzero value as the global fixed point changes abruptly (d, left).
    On the other hand, when the transition is of second order, the order parameter changes continuously to nonzero values (d, middle).
    }
    \label{fig:intro}
\end{figure*}

Despite its successes, compartmental modeling comes with some limitations. For instance, the aforementioned next-generation matrix formalism, while robust, offers a limited scope in addressing the multitude of degrees of freedom present in large-scale interconnected systems. In fact, the presence of this constraint requires the introduction of simplifying assumptions and approximations, often in the form of deterministic mean-field equations, to deal with the complexity inherent in these systems. Nevertheless, these approximations can still be very powerful. For example, in the case of complex networks with a heterogeneous degree distribution, it is possible to consider classes of a specific degree and then study the dynamics of the overall system by means of the dynamics of such classes. This approach leads to heterogeneous mean-field equations that allow one to model the collective behavior of a system more accurately than simple homogeneous mean-field equations~\cite{colizza2007reaction,colizza2008epidemic,gomez2011nonperturbative}.

In fact, this approach shares commonalities with the study of the broad class of reaction-diffusion processes to describe chemical reaction networks~\cite{unsleber2020exploration}, where units are chemical species undergoing transitions due to chemical reactions or spontaneous transformations. This parallel is striking, and it is, therefore, natural to wonder if, and under which conditions, it is possible to analyze population dynamics of a variety of complex systems by means of novel formalisms that overcome the inherent limitations of the current ones. 

Chemical reaction networks, in general, have been extensively studied using the Doi-Peliti formalism~\cite{Many,PELITI}, which offers a rich methodology to analyze stochastic and deterministic dynamics beyond pure mean-field approaches. However, while chemical reaction networks are subject to strict constraints, such as the mass balance of chemical elements~\cite{elser2000biological,sterner2017ecological}, this is not always the case for the general systems in which we are interested in this work, where compartments can also be of heterogeneous nature. Therefore, we anticipate that it will be necessary to adapt the Doi-Peliti formalism to deal with the variety of cases of interest for network and complexity science. Accordingly, in the following, we will refer, in general, to chemical reaction networks (CRNs) to indicate such a broad spectrum of systems, from chemical to ecological and social ones.

The dynamics of a CRN can be completely described by a master equation, and the Doi-Peliti formalism recalls the second quantization used in quantum field theory to apply the methods for studying quantum systems for the analysis of stochastic classical systems. One of these methods is the path integral representation~\cite{weber2017master}, that is particularly useful to go beyond the deterministic mean-field limit of a model, providing insight into its stochastic behavior. In the past years, attempts to extend this formalism to systems with many degrees of freedom have been proposed~\cite{phasetrans, Black, weber2017master, giuli2022dynamical}, although providing general results remains challenging. 

Building upon these works, we develop a framework that describes multicompartment systems by reducing them to a one-dimensional effective model, showing how to map a CRN to a Hamiltonian problem (see Fig.~\ref{fig:intro}). This mapping allows us to characterize the critical behavior of the CRN through an elegant geometrical interpretation. Furthermore, irrespectively of the presence, or lack thereof, of phase transitions in the stochastic process of interest, our formalism provides a way to approximate its steady state with high accuracy even for systems where existing approaches cannot be readily used. The idea of building on methods originating from quantum physics to study complex networks, or vice versa, has proven successful in stretching the boundaries of statistical physics~\cite{acin2007entanglement, paparo2012google, faccin2013degree, faccin2014community, de2015structural, de2016spectral, meng2021concurrence, villegas2023laplacian, ghavasieh2024diversity, rojas2024quantum}.
Similarly, we foresee broad applications of our dimensionality reduction approach to study the critical behavior and the population dynamics of multicompartment processes.

Before proceeding with the theoretical background, let us take the opportunity here to summarize the main steps to characterize the phase transition of generic multicompartment systems.
A visual representation of the procedure is shown in~\cref{fig:intro}.
First, we map a generic CRN into a Hamiltonian problem. 
The corresponding Hamiltonian can be obtained directly from the master equation of the original problem through the Doi-Peliti formalism (\cref{fig:intro:a,fig:intro:b}, \cref{subsec:compartment,subsec:path_integral}).
The second step is to perform an effective dimension reduction of the original problem by defining a reduced Hamiltonian through a proper parametrization (\cref{fig:intro:c}). 
To facilitate the analysis of the phase portrait, we apply a transformation that yields new variables with a clear physical interpretation (\cref{sec:theoretical_reduction}).
Finally, we identify critical points within the phase portrait as locations where multiple zero-energy lines of the reduced Hamiltonian intersect, signaling qualitative changes in the system's dynamics, i.e.~phase transitions.
Then, the order of the phase transition(s) can be determined by investigating the zero-energy lines as function of the control parameter (Fig.~\ref{fig:intro:d} and Sect.~\ref{sec:PhPortraits_PhTransitions}).
A schematic overview of these steps, with additional details, is presented in~\cref{tab:cookbook}. 
We stress that readers interested in the technical details continue to~\cref{sec:theoretical_background}, while those interested in the applications can proceed directly to~\cref{sec:results}.

%%%
\section{Theoretical background}\label{sec:theoretical_background}
\subsection{Stochastic compartmental models}\label{subsec:compartment} 
Many natural, biological and sociotechnical phenomena are stochastic processes. Through a probabilistic representation, one focuses on a macroscopic quantity $n$ whose dynamics is described by a distribution $p(n,t)$ that depends on the features of the system. Some examples include the spread of disease, where $n$ represents the number of infected individuals, or ecological systems, where $n$ describes species abundance. From a mathematical point of view, $n$ is the output of a discrete random variable, whereas $p(n,t)$ is the probability of finding that random variable with value $n$ at time $t$. Generally, we can take $n$ as a discrete quantity. We will focus on homogeneous Markovian systems only and, therefore, the dynamical behavior of the systems is governed by a master equation
\begin{equation}\label{eq:ME}
    \partial_t p(n,t)=\sum_{m}H_{nm}p(m,t),
\end{equation}
where the sum is evaluated over all possible states and $H_{nm}$ are the elements of a matrix defined by the system interactions. This matrix can be written in terms of the transition rates $\omega(n|m)$ that represent the probability per unit of time of transitioning from the state $m$ to the state $n\not= m$, that is,
\begin{equation}\label{eq:gen0}
 H_{nm} = \omega(n|m)-\delta_{nm}\sum_{m'}\omega(m'|n),
\end{equation}
where $\delta_{nm}$ is the Kronecker delta and $\omega(n|n)=0$.
This compartmental representation is very flexible, and the evolution of the system narrows down to keeping trace of the number of elements that belong to each state-compartment. 

Now consider $M$ to be the number of compartments necessary to fully characterize a system. In this case, $\boldsymbol{n} = (n_1,\ldots,n_M)$ is the (multidimensional) variable of interest, with $n_i$ being the population, abundance, or occupation number of its corresponding compartment $i$. 
For a generic system, $M$ can be arbitrarily large, and the interaction network very complex.
For the sake of simplicity, in this work we limit ourselves to the case of well-mixed chemical reaction networks, given by (a set of) transitions of the form
\begin{equation}\label{eq:CRN}
    \sum_{i=1}^M k_i A_i \xrightarrow{\lambda} \sum_{i=1}^M l_i B_i,
\end{equation}
reactions where $k_i$ and $l_i$ represent the stoichiometric coefficients for the chemical reactions~\cite{giuli2022dynamical}, and $\lambda$ represents the rate of a specific reaction.
Note that these are mass-action reactions, which means that the rates are proportional to the population numbers.

In general, studying the time evolution and the critical behavior of generic CRNs in the form of~\cref{eq:CRN} is difficult.
However, in what follows, we illustrate that our proposed formalism will simplify such analyses by focusing on the effective dynamics of a few compartments and ``freezing'' all the others (\cref{fig:intro}).
In its essence, we first employ the Doi-Peliti formalism and its corresponding path integral representation.
Then, we take a Wentzel-Kramers-Brillouin (WKB) approximation in order to obtain a set of Hamilton's equations, and we show that one can freeze a specific set of compartments to obtain the reduced dynamics.
This scheme effectively reduces the original problem of dimension $M$ to a problem of dimension $2M-2m$, where $m$ is the number of frozen compartments.
It, furthermore, allows us to study the phase transition geometrically, by looking at the intersections of the zero-energy lines of the Hamiltonian in a phase portrait.
These intersections are the fixed points of the reduced system, and we can readily identify the order of the phase transition and the location of the critical point. 
In certain cases, we are additionally able to approximate with high accuracy the distribution of the order parameter at stationarity, thus readily obtaining its fluctuations.
As such, our proposed formalism provides more information on the system than a mean-field approach can offer.
An overview of the dimension reduction scheme and the characterization of the phase transition(s) is depicted in~\cref{fig:intro}.

\subsection{The Doi-Peliti formalism and the path integral representation}\label{subsec:path_integral}%
The Doi-Peliti (DP) formalism has been introduced as a means to tackle generic CRNs of the form provided above (see, e.g., \citep{Many, PELITI}). 
Briefly, the formalism describes the dynamics of a compartment using creation and annihilation operators, which are denoted by $a^{\dagger}$ and $a$, respectively. 
Hence, analogously to the second quantization, an element of the system can be created or destroyed as defined by the transition rules encoded in the master equation.
This approach has been extensively covered in the literature (for detailed reviews, see~Refs.~\cite{weber2017master,delrazo2024field}). 
Note that we adhere to the standard definitions employed by others (e.g., \citep{PELITI, weber2017master,giuli2022dynamical, twolangevin}).
To this end, let us first develop the usual path integral representation.

Let us introduce the bra-ket notation to describe the state of a system, i.e.,
\begin{equation}
    \ket{\textit{\textbf{n}}}=\ket{n_1,\ldots,n_M}.
\end{equation}
The corresponding bra is given by the normalization condition
\begin{equation}\label{eq:normalization}
\bra{\textbf{\textit{n}}}\ket{\textbf{\textit{n}}'}=\prod_{i=1}^{M} n_i!\delta_{n_in_i'}
\end{equation}
With this normalization condition, the creating and annihilation operators are defined, respectively, by
\begin{equation}
    \label{eq:operators}   
    a_{i}^{\dagger}\ket{\textit{\textbf{n}}}=\ket{\textit{\textbf{n}}+\mathbb{I}_i}\,,\qquad a_i\ket{\textit{\textbf{n}}}=n_i\ket{\textit{\textbf{n}}-\mathbb{I}_i},
\end{equation}
where $\mathbb{I}_i$ is the canonical vector with the $i$-th component equal to 1 and all others zero. 
Thus, within the context of compartmental models, $a_i^{\dagger}$ and $a_i$, respectively, create and annihilate an element belonging to compartment $i$, and follow the usual bosonic commutation relations
\begin{equation}
    [a_i^\dagger,a_j^\dagger] = [a_i,a_j] = 0, \quad [a_i,a_j^\dagger] = \delta_{ij},
\end{equation}
where the commutator is defined as $\left[ a, b \right] = ab - ba$.
To obtain a path integral representation, let us introduce the generating function
\begin{equation}
    \label{eq:generating}
    \ket{G} = \sum_{\boldsymbol{n}} p(\boldsymbol{n},t)\ket{\boldsymbol{n}}, 
\end{equation}
and, by using~\cref{eq:normalization}, the probability distribution of attaining a specific state at time $t$ reads
\begin{equation}
p(\textit{\textbf{n}},t)=\frac{\bra{\textit{\textbf{n}}}\ket{G}}{\prod_in_i!}.
\end{equation}
Thus, $\ket{G}$ fully determines the statistical behavior of the system. By employing the master equation of~\cref{eq:ME}, we obtain a ``Schr\"odinger-like'' equation within the Doi-Peliti representation, which reads
\begin{equation}
    \label{eq:newME}
    \partial_t\ket{G}=\mathcal{H}\ket{G},
\end{equation}
where
\begin{equation}\label{eq:DPH}
    \mathcal{H}=\sum_{\bm n,\bm m}\frac{1}{\prod_im_i!}\ket{\bm n}H_{nm}\bra{\bm m}
\end{equation}
is the \emph{Doi-Peliti Hamiltonian} (DPH). 
For a generic CRN, $\mathcal{H}$ can be written as a polynomial function of the creation and annihilation operators (\cref{sec:DPHforCRN_app}). 

To have an explicit expression of $\ket{G}$, it is necessary to choose a basis for the Fock space.
We work with
\begin{equation}\label{eq:basis}
    \ket{\textit{\textbf{n}}}=\prod_{i=1}^Mx_i^{n_i},
\end{equation}
where $\vb*{x} = (x_1, \ldots, x_M)$.
Using the normalization condition, we have that 
\begin{align}
    \bra{\bm n} = \left[\prod\left( \pdv{{\,}}{x_i}\right)^{n_i}\right]_{\boldsymbol{x}=0}
\end{align}
Under these bases, the path integral representation for the generating function is given by~\citep{weber2017master}
\begin{equation}\label{eq:pathintegral1]}
   \ket{G} = \int_{[0}^{t)} \mathcal{D}[ \textbf{\textit{x}},   \textbf{\textit{q}}]\,e^{-\mathcal{S}[ \textbf{\textit{x}}, \iu \textbf{\textit{q}}] + 
   \sum_j{n_{0j}} \log{x_{0j}}}, 
\end{equation}
where $\iu$ is the imaginary unit, $\textbf{\textit{n}}_0$ is the initial state of the stochastic process, $\mathcal{D}[\textbf{\textit{x}},  \textbf{\textit{q}}]$ is a suitable measure~\citep{weber2017master, twolangevin}, and $\mathcal{S}$ is the action of the system defined as 
\begin{equation}\label{eq:action}
    \mathcal{S}[\bm{x}, \iu \bm{q}] = -\int_{0}^t \diff{\tau}
    \left[
    \sum_j \iu q_j \partial_{\tau}x_j + 
    \mathcal{H}( \bm x, \iu \bm q)
    \right],
\end{equation}
where $x_j = x_j(\tau)$ and $q_j = q_j(\tau)$.
Further details on the derivation of the generating function and on the notation are reported in~\cref{sec:PI_derivation_app}.

Within the path integral representation, the operators $a_j^{\dagger}$ and $a_j$ are replaced by two fields $x_j(\tau)$ and $\iu q_j(\tau)$. 
They are linked to the occupation numbers $n_j$ through the formal relation 
\begin{equation}
    \label{eq:formal_relation}
    \iu q_j x_j=n_j,
\end{equation}
which is derived from the definition of the number operators $\hat N_j=a_j^{\dagger}a_j$, where $\hat N_j\ket{ \textit{\textbf{n}}}=n_j\ket{ \textit{\textbf{n}}}$. 
Note that, despite handling~\cref{eq:pathintegral1]} being generally difficult, we show that in the case of chemical reactions we can use an approximation similar to the WKB one.
Doing this reduces the study of a stochastic process to a Hamiltonian problem.

\subsection{WKB approximation and stationary paths}\label{subsec:wkb}%
The path integral representation is formally exact, but, to obtain explicit results, one has to resort to approximations. 
Recently, it has been shown that the WKB approximation for the study of stochastic systems is a good candidate~\citep{Black, artime2018first,Assaf,WKBmix}. 
In~\cref{sub:wkb_app}, we show that this approximation can be applied to CRNs such as those in~\cref{eq:CRN}, thus greatly simplifying the path integral representation. 
In brief, this approach is valid when the system size (e.g., the total population) $N\gg 1$ and when the transition rates between $k$ reagents scale as $N^{k-1}$.

When these assumptions hold, the generating function $\ket{G}$ can be approximated as
\begin{equation}\label{eq:stationaryS}
    \ket{G} \approx \exp ( -\widetilde{\mathcal{S}} ),
\end{equation}
where $\widetilde{\mathcal{S}}$ is the action $\mathcal{S}$ evaluated over the solutions $(\widetilde{\vb*{x}},\widetilde{\vb*{q}})$ of the Hamilton equations 
\begin{align}
    \label{eq:Hsist}
    \pdv{\widetilde{\vb*{x}}}{t} = - \pdv{\mathcal{H}}{\widetilde{\vb*{q}},}, \qquad 
    \pdv{\widetilde{\vb*{q}}}{t} = \pdv{\mathcal{H}}{\widetilde{\vb*{x}}},
\end{align}
and $\mathcal{H}$ is the Doi-Peliti 
Hamiltonian (\cref{eq:DPH}). 
Although this system of equations might appear complicated, it actually represents a reduced description of the master equation and, as we show, provides valuable information about the dynamical and critical properties of the stochastic process.

As we work with chemical reaction networks where the WKB approximation applies, we consider only stationary trajectories, i.e., those in which the system size $N$ and the time $t$ are sufficiently large. 
In Appendix~\ref{sec:basis_app}, we show that $(\widetilde{\bm{x}},\widetilde{\bm{q}})$ can be taken as real functions satisfying the important relation
\begin{equation}
    \label{eq:nxq}
    n_j = \widetilde{x}_j \widetilde{q}_j.
\end{equation}
Note that this relation differs from Eq.~\eqref{eq:formal_relation}. Indeed, as is discussed in detail in Appendix~\ref{sec:basis_app}, we find that by applying the WKB approximation and changing the Fock space basis, we can substitute $\iu \boldsymbol{q}$ with a real field $\widetilde{\boldsymbol{q}}$, thereby providing a more intuitive description.

It is possible to show that the trajectory $\widetilde{x}_j=1$ is a solution of~\cref{eq:Hsist} (\cref{sec:DPHforCRN_app}).
This trajectory plays an important role from a statistical point of view, as the average value of $n_j$ is given by
\begin{equation}
    \label{eq:meanfielddynamics}
    \langle n_j\rangle=\pdv{\ket{G}}{x_j} \biggr\rvert_{\boldsymbol{x}=1} \equiv  \widetilde{q}_j,
\end{equation}
where we use~\cref{eq:stationaryS}. 
Thus, in the phase space, the deterministic behavior of the system --- that is to say, the \emph{mean-field behavior} --- is contained on the \emph{mean-field manifold} $\widetilde{\boldsymbol{x}}=1$. Put otherwise, $\widetilde{\boldsymbol{x}}=1$ contains all deterministic (mean-field) trajectories.
As we illustrate in~\cref{sec:PhPortraits_PhTransitions}, this manifold is key in the analysis of the critical behavior of CRNs.

To provide a more intuitive physical interpretation of our variables, we employ the Cole-Hopf (CH) transformation~\citep{CH}, which reads
\begin{align}
    \label{eq:colehopf}
    x_j = e^{\iu \theta_j}, \qquad \iu q_j = \eta_j e^{-\iu \theta_j}.
\end{align}
We stress that this transformation is a formal procedure, since there are no real variables that satisfy such relation. 
Yet, in doing so, we replace the field $(\boldsymbol{x}, \iu \boldsymbol{q})$ by $(\boldsymbol{\eta}, \mathrm{i} \boldsymbol{\theta})$, where $\eta_j$ represents the population number $n_j$.
With this transformation in place, the action $\mathcal{S}$ now becomes 
\begin{equation}
    \mathcal{S} = \int_{t_0}^t \diff \tau \left[ 
        \sum_j \iu \theta_j \partial_\tau \eta_j - 
        \widehat{\mathcal{H}}
    \right] + \iu \boldsymbol{\theta}_0 \boldsymbol{\eta}_0 - 
    \iu\boldsymbol{\theta}\boldsymbol{\eta}
\end{equation}
where $\widehat{\mathcal{H}}$ is the Hamiltonian after applying the Cole-Hopf transformation.
As was done for $(\bm x,\iu \bm q)$, in the stationary limit we can replace $(\bm \eta,\iu \bm\theta)$ by the (real) solutions of
\begin{equation}
    \label{eq:h2sist}
    \pdv{\widetilde{\boldsymbol{\eta}}}{t} =  \pdv{\widehat{\mathcal{H}}}{\widetilde{\boldsymbol{\theta}}}, \qquad
    \pdv{\widetilde{\boldsymbol{\theta}}}{t} = -\pdv{\widehat{\mathcal{H}}}{\widetilde{\boldsymbol{\eta}}},
\end{equation}
which is the Hamiltonian system of~\cref{eq:Hsist} rewritten using the Cole-Hopf variables.
With this change of coordinates, the mean-field dynamics is recovered on $\widetilde{\boldsymbol{\theta}}=0$. 
Note that the mean-field manifold $\widetilde{\bm{x}}=1$ and $\widetilde{\bm{\theta}}=0$ are \emph{always} solutions of~\cref{eq:Hsist,eq:h2sist}, respectively.
Therefore, this manifold is a universal feature of CRNs~\cite{statpath} (see also~\cref{sec:DPHforCRN_app}).
As such, the physical interpretation of $\bm{\theta}$ relates to the degree in which the dynamical behavior of the system is close to the deterministic one.
That is, $\bm{\theta}$ expresses the role of stochastic fluctuations \emph{beyond} the mean-field approximation. 
Therefore, by studying the phase portrait of Eqs.~\eqref{eq:h2sist}, we can investigate the time evolution by means of its attractors.
Moreover, we show that we can probe the role of fluctuations (\cref{sec:reduced}).

As we work with systems wherein we can use the WKB approximation (see~\cref{sec:basis_app}), we can replace the fields $(\boldsymbol{x},\textrm{i}\boldsymbol{q})$ with the real fields $(\widetilde{\boldsymbol{x}},\widetilde{\boldsymbol{q}})$.
Going forward, to simplify notation, we drop the tilde and simply use $(\boldsymbol{x},\boldsymbol{q})$ to denote the real fields that are the stationary solutions of~\cref{eq:Hsist} (and $(\boldsymbol{\eta},\boldsymbol{\theta})$ their corresponding Cole-Hopf variables admitting~\cref{,eq:h2sist}), such that now we have $n_j = x_j q_j$. 
In addition, in the following sections, we freely alternate between the $(\boldsymbol{x},\boldsymbol{q})$ representation and the Cole-Hopf $(\boldsymbol{\eta}, \boldsymbol{\theta})$ one.
Generally, the former tends to be more suitable for analytic computations, while the latter is often more convenient for visualizing the phase portraits and for investigating the critical behavior.

\subsection{Characterization of phase transitions under general constraints}
\label{sec:PhPortraits_PhTransitions}%
Following the ideas put forward in Ref.~\citep{phasetrans}, we build on the formalism presented so far to develop a mathematical framework for the analysis of phase transitions. 
Using our proposed formalism, we obtain a strategy for studying the phase transitions of generic chemical reaction networks. 

We first write the Doi-Peliti Hamiltonian of the system and
use it to find the fixed points on the zero-energy sets. 
When looking for the zero-energy sets, it is essential to account for all possible constraints.
Within the context of CRNs, such constraints typically are related to the conservation of the total number of elements.
After identifying and imposing the constraints, the dimension of the system will be reduced to $2M-2m$, where $m$ is the number of constraints, yielding a reduced Hamiltonian $\mathcal{H}^\star$. In~\cref{sec:theoretical_reduction}, we provide more details on how to select the constraints and build the reduced Hamiltonian.
Then, we consider the simplified dynamical system  (\cref{eq:Hsist}) with $\mathcal{H}^\star$, for which we define the Jacobian relative to the reduced system as
\begin{equation}
    \label{eq:jacobian}
    \mathbf{J}=
    \begin{pmatrix}
    -\dfrac{\partial \mathcal{H}^\star}{\partial \boldsymbol{x} \partial\boldsymbol{q}} & 
    -\dfrac{\partial \mathcal{H}^\star}{\partial \boldsymbol{q} \partial\boldsymbol{q}} \\[2.5ex]
    \dfrac{\partial \mathcal{H}^\star}{\partial \boldsymbol{x} \partial\boldsymbol{x}} & 
    \dfrac{\partial \mathcal{H}^\star}{\partial \boldsymbol{x} \partial\boldsymbol{q}}
    \end{pmatrix}
\end{equation}
We then look for the conditions that make the determinant of $\mathbf{J}$ null, which occurs when $\mathbf{J}$ has at least one null eigenvalue.
Finally, we look for the parameter relations that characterize the phase transition. 
We assume that our system always converges to a stationary state, and we describe it by means of a nonequilibrium order parameter $\varphi$, which denotes the asymptotic density of a specific compartment. 
If we take, without loss of generality, $n_1 \approx N$, then
\begin{equation}
    \label{eq:rho}
    \varphi = 1-\lim_{t\to\infty} \frac{\langle n_1\rangle}{N}.
\end{equation}
For instance, in the context of an epidemiological CRN, we would assume that, when an outbreak of a new disease is about to occur, the number of susceptible individuals in a population is very large, $n_S \approx N$. Then, our framework allows us to study the dynamics and the critical behavior of the system when a small number of infected individuals is introduced in the populations.
By varying the (control) parameters near the critical point(s), we can determine whether there is a first- or second-order phase transition, as we describe below.
As we are interested in the large time behavior, we consider only the null eigenvalue of the Hamiltonian system; that is, we focus our attention on the zero-energy sets where 
\begin{equation}
    \label{eq:zeroenergy}
    \mathcal{H}^\star = 0 \qquad(\textrm{or}\quad 
    \widehat{\mathcal{H}}^\star = 0)
\end{equation}
Within the WKB approximation regime, we can study the topology of these sets to obtain information about the system at large times. 
Here, by topology, we mean the trajectories on the phase portrait related to $\mathcal{H}^\star$. 
The fixed points of the Hamiltonian system on the level set $\mathcal{H}^\star=0$ play an important role, since they are related to the stable and the unstable manifolds that determine the evolution of the system~\citep{dyna}.
In particular, the dynamics converges at the hyperbolic fixed point on the stable manifold, and diverges from it on the unstable one. 
Therefore, when modifying the control parameters, such as the transition rates, the configuration of the fixed points can change.
As a result, the dynamics of the system will strongly vary as its attractors are modified, suggesting the presence of phase transitions.
Recall that $\bm{x}=1$ is the mean-field manifold for which $\mathcal{H}^\star=0$ and, thus, represents the mean-field dynamics (see~\cref{eq:meanfielddynamics}). 
By studying the fixed points on this manifold, we recover the mean-field phase transitions.
Moreover, by analyzing the topology of the stable and unstable manifold, we are able to characterize these transitions~\citep{phasetrans} (see also~\cref{fig:intro} and~\cref{sec:results}).
If new global attractors suddenly appear when the control parameters are modified, then the dynamics of the system radically changes and, thus, the system admits a first-order phase transition. 
If new attractors do not appear and the positions of the existing ones change continuously, the system instead displays a second-order phase transition.

We stress that we consider systems that have absorbing states; hence, they are out of equilibrium and cannot be treated using the standard tools of equilibrium statistical mechanics.
The nonequilibrium properties, however, can be addressed by leveraging the Doi-Peliti mapping to the Hamiltonian representation.

The considered constraints that helped us reduce the system dimension need to be taken into account in the description of the phase transitions. We represent these by an equation of the form $\bm{C} = \mathcal{C}(\bm{x},\bm{q})$, where $\bm{C}$ is now an $m$-dimensional vector. Assuming that the implicit function theorem holds, we can find a function $Q$ such that, for $m$ components of $\bm{q}$, we have
\begin{equation}
\label{eq:vincoli}
    (q_1,\ldots,q_m)=Q(q_{m+1},\ldots,q_M,x_1,\ldots,x_M,\bm{C})
\end{equation}
Furthermore, constraints represent first integrals of the Hamiltonian system. 
By substituting~\cref{eq:vincoli} into $\mathcal{H}$, we obtain a Hamiltonian that is independent of these variables; therefore, the time derivatives of the corresponding conjugate coordinates $(x_1,\ldots,x_m)$ are all zero, and, thus, we can let $x_i=1$ for all $i = 1, \ldots, m$. 
Each constraint reduces the dimension of the system by $2$, projecting the DPH onto a new, reduced Hamiltonian $\mathcal{H}^{\star}$ that depends on $2M-2m$ variables. Under these premises, the system reads
\begin{align}
    \label{sist:crit}
    \partial_{x_j} \mathcal{H}^\star = 0, \qquad \partial_{q_j} \mathcal{H}^\star = 0,
\end{align}
with the constraints 
\begin{align}
    \label{eq:constr}
    x_j = 1, \qquad \det (\mathbf{J}) = 0,
\end{align}
where $\textbf{J}$ the Jacobian depends on $\bm x$ and $\bm q$ (see~\cref{eq:jacobian}). 
By treating the coefficients of the transition rates as variables, the system of~\cref{sist:crit} gives us the critical points for possible phase transitions. 
In the same way, we can build this system by using the Cole-Hopf coordinates. A more technical discussion of how to use a Hamiltonian representation to study CRN phase transitions, including conservation quantities and bias, is explored in Ref.~\cite{lazarescu2019}.

The approach presented so far is a generalization of the results discussed in Ref.~\cite{phasetrans} to multicompartment systems that include $m$ constraints. In order to exploit these results and apply the framework to arbitrary multidimensional CRNs, we first need to find a way to systematically identify these constraints. In the next section, we present our novel method for finding $m=M-1$ conservation laws, thus reducing the system's dimension to a single effective degree of freedom that focuses solely on the order parameter. %variables.

\subsection{Dimension reduction}\label{sec:theoretical_reduction}%
Dynamical models with many components are common, and their characterization as Hamiltonian systems can be complicated. 
Moreover, only in the case of single-compartment systems, are we able to find a graphical visualization of the level sets with the tools presented so far. 
In the previous section, we have observed that the presence of first integrals helps simplify multicompartment systems. 
We employ this property to build our own dimension reduction method, which is based on using the deterministic equations as first integrals to
effectively ``freeze" the degrees of freedom that are \emph{not} inherent to the order parameter (\cref{fig:intro}). 
Within this context, freezing a variable (or compartment) is equivalent to saying that one constrains this variable to live on the mean-field manifold where $x_j=1$ (or $\theta_j=0$; see~\cref{eq:meanfielddynamics}). Using this approach, we obtain an \emph{effective}, or \emph{reduced}, Hamiltonian $\mathcal{H}^{\star}$ that captures the fundamental features of the system. 

To show this, let us initially  consider a Hamiltonian $\mathcal{H}$ of a general process. 
The deterministic trajectories are
\begin{equation}
    \label{eq:determtrajectories}
    \partial_t \langle n_i \rangle = 
    f_i \big( \langle n_1 \rangle, \ldots, \langle n_M \rangle \big),
\end{equation}
where $f_i$ are functions of the occupation numbers obtained as the solutions of the Hamilton's equations by setting $x_j=1$ and $q_j=\langle n_j\rangle$  for all $j\neq i$. 
Since we assume that a steady state is asymptotically reached when $t\to\infty$, this set of equations admits the solution for $\partial_t\langle n_j\rangle=0$. 
Without loss of generality, we focus on the first compartment. 
Then, for large $t$, we rewrite the previous system as
\begin{subequations}
\label{eq:integraliprimi}
\begin{align}
    \partial_{t}\langle n_1 \rangle &= f_1(\langle n_1\rangle, \ldots, \langle n_M \rangle) \\
    \partial_{t}\langle n_j \rangle &= 0
\end{align}
\end{subequations}
meaning that now $f_j(\langle n_1 \rangle, \ldots, \langle n_M \rangle)=0$ for all $j>1$.
That is to say, all compartments but the first (i.e., the one related to the order parameter) are frozen, again meaning that $x_j=1$ (see above, and see~\cref{fig:intro,eq:meanfielddynamics}). In doing so, we are imposing constraints on the average values of the occupation numbers, from which we derive the set of first integrals. We stress again that imposing these constraints is equivalent to saying that all compartments but the one related to the order parameter are confined to the mean-field manifold where $x_j=1$. 
Hence, the dynamics of all these compartments can be described as a function of $n_1$ and,
therefore, letting $n_1$ converge to its steady state means that all other compartments simultaneously converge to their steady state as well.

Further physical insights can be gained by studying what occurs near this limit. Let $x_1$ and $q_1$ be the only variables related to the order parameter, while all the other degrees of freedom are frozen. From an analytic point of view, using Eqs.~\eqref{eq:integraliprimi} we can rewrite $q_{j}$ for all $j>1$ as functions of $q_1$ and $x_1$. 
Although this is not always possible, we show that we can add a suitable transition rate to make Eqs.~\eqref{eq:integraliprimi} solvable without altering the critical environment (see~Appendix~\ref{sec:closure_app} for more details). 
In other words, we can define a parametrization $\Lambda$ such that
\begin{align}
    \label{eq:Lambda}
    \Lambda(x_1, q_1) &= 
    \left\{
    x_1, x_j = 1, 
    q_1, q_j = Q_j(x_1,q_1)
    \right\},
\end{align}
where $j>1$, and recall that $x_j=1$ and $q_j = Q_j(x_1,q_1)$ with $Q_j$ as defined in~\cref{eq:vincoli}.
By applying this parametrization to the original DPH, we obtain the reduced Hamiltonian,
which, thus, reads
\begin{equation}
    \label{eq:reducedhamiltonian}
    \mathcal{H}^{\star} = \mathcal{H}\circ\Lambda   
\end{equation}
where $\circ$ is the composition operator, i.e., we have
\begin{equation}
    \mathcal{H}^\star = 
    \mathcal{H}\big(x_1,x_j=1,q_1,q_j=Q_j(x_1,q_1)\big)
\end{equation}
and, thus, $\mathcal{H}^\star$ depends only on $x_1$ and $q_1$.
Since the generating function is defined in terms of the $(x,q)$-coordinates, it is natural to employ this notation to represent the parametrizations and the reduced Hamiltonian. We then proceed with the Cole-Hopf transformation to obtain a more transparent physical interpretation.

In some cases, the solutions of~\cref{eq:integraliprimi} can yield different parametrizations. 
These may represent different states of the system, where the corresponding reduced Hamiltonians have different hyperbolic fixed points.  
To construct the correct Hamiltonian, we can investigate the stability of the fixed points of the full problem, and choose the parametrization such that $\mathcal{H}^{\star}$ has the same attractor(s) as the complete system. More specifically, we select as a \emph{domain} of the parametrization $\Lambda$ the one corresponding to the set of parameters for which the equilibria of the full system are recovered. If, for a point in the parameter space, two parametrizations yield the same equilibrium, then this point is generally a critical point.
In fact, if varying the system's parameters leads to the same stationary states for different parametrizations, it is implied that there is a superposition of distinct fixed points.
As previously discussed in~\cref{sec:PhPortraits_PhTransitions}, this indicates that a phase transition occurs. Since for many systems we have a number $m$ of conservation laws, the dimension of the problem is reduced by $2m$.
We have extra degrees of freedom, since we can derive different parametrizations $\Lambda$ depending on the first integrals that we are choosing when deriving an effective one-dimensional system. This selection is not entirely arbitrary. In fact, we must ensure that the set of stationary equations and the conservation laws admit a non-null solution (for a more detailed discussion, see~\cref{sec:spatiallotkavolterra}).
Once the unusable stationary equations are discarded, we find a degeneration of possible parametrizations and reduced Hamiltonians. 

Even if the dimension reduction of some CRNs admits several parametrizations, we highlight that this fact does \emph{not} change the deterministic description. Despite the multiplicity of reduced Hamiltonians $\mathcal{H}^{\star}$, if the corresponding $\Lambda$ satisfies the condition discussed above, the fixed points and the asymptotic behavior of the system do not vary. 
From a stochastic point of view (see also~\cref{sec:reduced} below), we obtain different Langevin and Fokker-Planck equations for each parametrization.
This means that the stochastic information that we get is related to the conditional probability $p(n_1|n_2,\ldots,n_{M})$, which we are taking as an approximation for $p(n_1)$ in the stationary limit.
Different parametrizations lead to different constraints and, thus, to different conditional probabilities. 

Remarkably, once the parametrization degeneracy is overcome, our dimension reduction scheme proves to be a useful way to describe the stochastic behavior of a large variety of models. 
Using this, we are indeed able to properly characterize both the order parameter of the phase transition via the mean value of the population of a single compartment (\cref{eq:rho}), and its fluctuations.
As we discuss in the next section, under some assumptions, we are able to approximate the distribution $p(n_1)$ incredibly well, meaning that \emph{all} its moments are accessible. 
On the other hand, when the parametrization degeneracy cannot be settled, higher moments of the occupation number $n_1$ are not properly captured and additional assumptions must be made. The physical intuition behind this is that, when the dimensions are reduced, we fix not only the average but also the noise of all degrees of freedom other than the order parameter by setting $x_j=1$~(or, $\theta_j=0)$, and, thus, some information is unavoidably, and perhaps unexpectedly, lost.

Finally, we remark that the dimension reduction is not able to describe the transient dynamics of the systems, as it describes only its stationary limit. 
However, as we show, it remains very useful for the study of critical behaviors. 

\subsection{Fokker-Planck and Langevin representations}\label{sec:reduced}%

Now that we have illustrated our dimension reduction scheme, let us illustrate how it can be applied to obtain reduced Fokker-Planck and Langevin descriptions of the dynamics. 
We apply the largest possible dimension reduction $m=M-1$.
That is to say, we freeze all but a single compartment related to the order parameter.
Then, recall that the reduced Hamiltonian $\widehat{\mathcal{H}}^\star$ defines a Hamilton system in the Cole-Hopf variables. These variables allow us to represent the fluctuations around the mean-field dynamics by means of $\bm{\theta}$. 
We have seen that the first derivative of $\widehat{\mathcal{H}}^\star$ with respect to $\boldsymbol{\theta}$ evaluated on the mean-field manifold describes the mean-field dynamics of the reduced system. 
Next, we show that its second derivative allows us to capture the systems' fluctuations (see also Refs.~\citep{giuli2022dynamical, weber2017master}).

When the population number $n$ is sufficiently large, the reduced Hamiltonian $\widehat{\mathcal{H}}^{\star}$ in the Cole-Hopf representation can be approximated by a continuous variable.
In doing so, we can rewrite the master equation of~\cref{eq:ME} using a Kramers-Moyal expansion~\citep{weber2017master}:
\begin{align}
    \label{eq:KM}
    \partial_tp(n,t) &= \sum_{m\ge 1} \frac{(-1)^m}{m!}
    \pdv{^m}{n} \big[W_m(n)p(n,t)\big] \nonumber \\
    &= \mathcal{H}_{\textrm{KM}}(n,\partial_n)p(n,t)
\end{align}
with
\begin{equation}
    W_m(n)=\int_{\mathbb{R}} \diff{y}\,y^m\omega(n+y|n),
\end{equation}
and where $\mathcal{H}_{\textrm{KM}}$ is the Kramers-Moyal operator. 
In this limit, there is a link between $\widehat{\mathcal{H}}$ and $\mathcal{H}_{\textrm{KM}}$. Using the definition of $\mathcal{H}$ and considering~\cref{eq:DPH}, one can show that for a one-dimensional system (\cref{sec:DPHforCRN_app_subsecKM})
\begin{equation}
    \label{eq:re}
    \mathcal{H}^{\dagger}_{\textrm{KM}}(\eta, \theta ) = \widehat{\mathcal{H}}^\star(\eta,\theta),
\end{equation}
where $\mathcal{H}^{\dagger}_{\textrm{KM}}$ is the adjoint operator of $\mathcal{H}_{\textrm{KM}}$. Note that  under this mapping, we have $\eta=n$ and $\theta=\partial_{\eta}$. 

Interestingly, the same reasoning can be applied to systems with more degrees of freedom by using the dimension reduction presented above. If we consider the average $\langle n_1 \rangle = \eta$ as the order parameter, we can expand $\widehat{\mathcal{H}}^{\star}$ up to the second order of $\theta$ around $\theta=0$ to obtain a Fokker-Planck equation~\citep{vankamp}
\begin{equation}
    \label{eq:fokker_planckKM}
    \partial_t p(\eta,t) = -\partial_\eta \big[\partial_\theta \mathcal{H}^\star_0 p(\eta, t) \big]
    + \partial^2_\eta \big[\tfrac{1}{2} \partial^2_\theta \mathcal{H}^\star_0 p(\eta,t) \big],
\end{equation}
where, with some slight abuse of notation, we have $\partial^k_\theta\mathcal{H}^\star_0 = [\partial^k_\theta \widehat{\mathcal{H}}^\star]_{\theta=0}$ and $p(\eta,t)$ is the conditional probability distribution constrained by the stationary conditions. 
Hence, by expanding the reduced Hamiltonian, we can write an effective one-dimensional Fokker-Planck equation for the multicompartment stochastic process, which makes further calculations considerably easier. 
From the Fokker-Planck equation, we can readily obtain the Langevin equation of the problem~\citep{vankamp}
\begin{equation}
    \label{eq:efflangevin}
    \dv{\eta}{t} = v + \sqrt{w} \cdot \xi_t,
    %\dv{\eta}{t} &= \partial_\theta \mathcal{H}^\star_0 + 
    %\sqrt{\partial^2_\theta \mathcal{H}^\star_0}\cdot \xi_t,
\end{equation}
where we have defined the drift and noise strength 
\begin{equation}
    \label{eq:driftnoisestrength}
    v = \partial_\theta \mathcal{H}_0^\star, \qquad
    w = \partial_\theta^2 \mathcal{H}_0^\star,
\end{equation}
respectively, and $\xi_t\equiv\xi(t)$ Gaussian white noise in the It\^o representation. More details on these derivations are given in~\cref{sec:lang_appendix}.
Formal steady-state solutions of~\cref{eq:fokker_planckKM} are known to be~\cite{pdf}
\begin{equation}\label{eq:solFP}
     \lim_{t\to\infty}p(\eta,t)=\mathcal{Z} \,e^{-V(\eta)},
\end{equation}
with the potential
\begin{align}
     V(\eta)=\int \diff{\eta} \;
     \frac{
        \partial_{\eta}
        (\partial^2_\theta \mathcal{H}^\star_0)
        - 2 \partial_\theta \mathcal{H}^\star_0
     }{
     \partial^2_\theta \mathcal{H}^\star_0
     },
\end{align}
and $\mathcal{Z}$ a normalization factor. 
From this, it is clear that $\partial_{\theta}^2\mathcal{H}^{\star}$ is related to the system's fluctuations, and, by employing the Doi-Peliti formalism, we can have access to both the deterministic \emph{and} the stochastic descriptions.

It is important to note that~\cref{eq:solFP} is valid at equilibrium and represents an excellent approximation of the stationary distribution when $\eta$ is sufficiently far from the absorbing state (if any exist).
In systems with absorbing states, the stationary distribution collapses into a Dirac delta at the absorbing point. 
However, the transition rate to the absorbing state within the active regime scales as $e^{-N}$, with $N$ the system size (e.g., see~\cite{khasin2009extinction}, where $N$ is the total population for an epidemiological model).
Therefore, for intermediate times large enough to find stationary behavior but lower than the absorbing time, we can accurately approximate the stationary distribution with~\cref{eq:solFP}.

%%% COOKBOOK
\begin{table*}
\centering
\begin{NiceTabular}{l l l r}[colortbl-like]
  \CodeBefore
  \rowcolor{gray!50!white}{1}
  \rowcolor{gray!20!white}{2,7,13,18}
  \rowcolor{ylw}{4,6,9,11,15,20,22}%
  \Body
  \toprule
  \Block[l]{1-1}{\textbf{Steps}} && \Output{output},\;\Input{input}&
  \textbf{Refs.}
  \\
  \toprule
  % (Fig. 1a) &&& Section 1. \\
  \Block[l]{1-2}{1.~construct \textbf{Doi-Peliti Hamiltonian} $\mathcal{H}$}
  &&
  \Block[r]{1-2}{%
  \labelcref{subsec:compartment}, \labelcref{subsec:path_integral}
  } \\
  \emph{a.} &write master equation using the transition matrix $H_{nm}$ &
  $\partial_t \Output{p(\boldsymbol{n},t)} = \sum_{\boldsymbol{m}} \Input{H_{nm}} p(\boldsymbol{m},t)$ & \cref{eq:ME}\phantom{0} \\
  \emph{b.} &introduce creation and annihilation operators $a_i^\dagger$, $a_i$ &
  $a_i^\dagger\ket{\boldsymbol{n}} = \ket{\boldsymbol{n} + \mathbb{I}_i}$,\;
  $a_i\ket{\boldsymbol{n}} = n_i \ket{\boldsymbol{n} - \mathbb{I}_i}$ \phantom{M}
  & \cref{eq:operators}\phantom{0} \\
  \emph{c.} &compute the generating function $\ket{G}$ &
  $\Output{\ket{G}} = \sum_{\boldsymbol{n}} \Input{p(\boldsymbol{n},t)} \ket{\boldsymbol{n}}$ & \cref{eq:generating}\phantom{0} \\
  \emph{d.} &apply Doi-Peliti formalism to obtain the Hamiltonian $\mathcal{H}$ &
  $\Input{\partial_t \ket{G}} = \Output{\mathcal{H}} \ket{G}$ &
  \cref{eq:newME} \\
  \midrule%
  \Block[l]{1-2}{2.~compute \textbf{reduced Doi-Peliti Hamiltonian} $\mathcal{H}^*$} &&&
  \labelcref{sec:theoretical_reduction}
  \\
  \emph{a.}& identify order parameter\textsuperscript{1} $\varphi$ &
  $\Output{\varphi} = 1 - \langle \Input{n_i} \rangle / N$ &
  \cref{eq:rho} \\
  \emph{b.} &apply field transformations &
  $\Output{x_i} \Leftrightarrow \Input{a_i^\dagger}$,\;
  $\Output{q_i} \Leftrightarrow \Input{a_i}$ &
  \cref{eq:formal_relation} \\
  \emph{c.} &freeze all compartments but compartment $i$ & $x_j = 1$ for all $j \neq i$ &
  \cref{eq:constr} \\
  \emph{d.} &apply constraints\textsuperscript{2} and obtain parametrization $\Lambda$ &
  $\Output{\Lambda} = \{ \Input{x_i}, \Input{x_j=1}, \Input{q_i}, \Input{Q_j(x_i,q_i)} \}$ &
  \cref{eq:Lambda} \\
  \emph{e.} &apply $\Lambda$ to obtain the reduced Hamiltonian $\mathcal{H}^\star$ &
  $\Output{\mathcal{H}^\star} = \Input{\mathcal{H}} \circ \Input{\Lambda}$ &
  \cref{eq:reducedhamiltonian} \\
  \midrule%
  \Block[l]{1-2}{3.~analyze \textbf{phase portrait} and characterize \textbf{phase transitions}
  }
  &&&
  \labelcref{sec:PhPortraits_PhTransitions}
  \\
  \emph{a.}& apply Cole-Hopf transformations to obtain $\widehat{\mathcal{H}}^\star = \mathcal{H}^\star(\eta_i,\theta_i)$ \phantom{M}&
  $\Input{x_i} = e^{\Output{\theta_i}}$,\; $\Input{q_i} = \Output{\eta_i} e^{-\Output{\theta_i}}$ &
  \cref{eq:colehopf} \\
  \emph{b.}& find zero-energy lines of reduced Hamiltonian $\widehat{\mathcal{H}}^\star$ &
  $\widehat{\mathcal{H}}^\star = 0$ &
  \cref{eq:zeroenergy} \\
  \emph{c.\;i}& identify critical points where zero-energy lines cross & --- &
  \Block[r]{2-1}{\cref{fig:intro:d}} \\
  \phantom{c.\;}\emph{ii}& inspect critical points by changing control parameter(s) & --- &
  \\
  \midrule%
  \Block[l]{1-2}{4.~analyze \textbf{fluctuations} around steady state $\partial_t \eta_i=0$} &&&
  \labelcref{sec:reduced}  
  \\
  \emph{a.}& compute partial derivatives of $\widehat{\mathcal{H}}^\star$ at $\theta=0$ &
  $\Output{\partial_{\theta}^k \mathcal{H}_0^\star} = [ \partial_{\theta}^k \Input{\widehat{\mathcal{H}}^\star}]_{\theta=0}$ &
  --- \\
  \emph{b.}& obtain (reduced) drift $v$ and noise strength $w$ &
  $\Output{v} = \Input{\partial_{\theta} \mathcal{H}^\star_0}$,\;
  $\Output{w} = \Input{\partial_{\theta}^2 \mathcal{H}^\star_0}$ &
  \cref{eq:driftnoisestrength} \\
  \emph{c.}& write reduced Langevin equation &
  $\Output{\partial_t \eta_i} = \Input{v} + \sqrt{\Input{w}} \cdot \xi_t$ &
  \cref{eq:efflangevin} \\
  \bottomrule
\end{NiceTabular}
\label{Tab:tab1}
\caption{%
  \textbf{Methodological overview of our approach.}
  Detailed work flow with the steps one needs to take to characterize phase transitions and analyze fluctuations of generic multicompartment chemical reaction networks (for more technical details, see~\cref{sec:theoretical_background}).
  After constructing the Doi-Peliti Hamitonian (1, \cref{fig:intro:a,fig:intro:b}, \cref{sec:DPHforCRN_app}), we reduce the dimensions of the problem and obtain an effective, reduced, Hamiltonian $\mathcal{H}^\star$ (2, \cref{fig:intro:c}), which can be used to analyze the phase portrait (3, \cref{fig:intro:d}) and, under some conditions (see~\cref{sec:reduced} and \cref{sec:app:reduced}), the fluctuations around the steady state (4).
  \textsuperscript{1}While $\varphi = 1 - \langle n_i \rangle / N$ is common, other functional forms of the order parameter exist, and the form depends on the system one considers.
  \textsuperscript{2}A common constraint often applied to reduce the dimensions of the system is population conservation, i.e. $N = \sum_i n_i$.
  Here, we additionally use fixed points of the mean-field dynamics, i.e. $\partial_t \langle n_j \rangle = [\partial \mathcal{H}^\star / \partial x_j ]_{\boldsymbol{x}=1} = 0$ to obtain additional constraints (see~\cref{eq:integraliprimi}).
  Note that these constraints result from freezing all compartments but the one related to the order parameter.
}
\label{tab:cookbook}
\end{table*}

We note that the fluctuations experienced by the individual elements within the compartments of typical CRNs can be highly correlated. To see this, let us describe the (high-dimensional) stochastic dynamics as an It\^o stochastic differential equation, or chemical Langevin equation, as
\begin{equation}
    \label{eq:chemicallangevin}
    \partial_t \boldsymbol{n} = \mathbf{V} + \mathbf{W}^{1/2} \cdot \boldsymbol{\xi}_t
\end{equation}
 where $\mathbf{V}$ is the drift and $\mathbf{W}$ the correlation matrix.
When $\mathbf{W}$ is a diagonal matrix, each degree of freedom will have a single noise term in the corresponding Langevin equation. Hence, freezing the degrees of freedom is equivalent to replacing, in the Langevin dynamics of the order parameter, all stochastic variables but the nonfrozen one by their averages.
However, the correlation matrix  $\mathbf{W}$  may not be diagonal.
This occurs when CRNs contain spontaneous reactions.
These reactions give rise to fluctuations in a specific compartment that also depend on the occupation numbers of other compartments in addition to its own. 
For instance, in the spontaneous reaction $I \xrightarrow{\lambda} S$, fluctuations in $S$ are correlated with fluctuations in $I$.
As our reduced Langevin equation describes the noise as a (potentially complicated) second derivative of the reduced Hamiltonian on the mean-field manifold, $\partial^2_\theta \mathcal{H}^\star_0$, the stochastic dynamics of such CRNs may not be well approximated by the reduced Langevin equation of~\cref{eq:efflangevin}.
In contrast, when a CRN does not contain spontaneous reactions --- which typically occurs in systems where compartmental changes cannot occur, such as in ecological models like the (generalized) Lotka-Volterra model --- the dimension reduction scheme results in a reduced Langevin equation (\cref{eq:efflangevin})  that readily captures stochastic effects, and, hence, reliably provides information beyond the mean-field dynamics of a system.

%%% MODEL TABLE
\begin{table*}[t!]
    \begin{NiceTabular}{c c c c c c}[colortbl-like]
  \CodeBefore
  \rowcolor{gray!50!white}{1}
  \rowcolor{ylw}{3,5,7}
  %\rectanglecolor{red!30!white}{1-3}{8-3}
  \Body
  \toprule%
  \RowStyle[bold]{}
  model & compartments & reactions & order parameter & critical behavior & section \\
  \midrule
  $SIS$ & 2 & 2 & $1 - \langle n_S \rangle / N$ & second order &
  Sec.~\labelcref{subsec:SIS} \\
  modified $SIS$ & 2 & 3 & $1 - \langle n_S \rangle / N$ & tricrital point &
  Sec.~\labelcref{subsec:modifiedSIS} \\
  predator-prey\textsuperscript{a} & 3 & 4 & $\langle n_A \rangle$ & second order &
  Sec.~\labelcref{sec:spatiallotkavolterra} \\
  generalized Lotka-Volterra\textsuperscript{b} & 24 & 600 & $\langle n_1 \rangle$ & second order &
  Sec.~\labelcref{subsec:fluctuations} \\
  \midrule%
  tax evasion & 3 & 5 & $1 - \langle n_H \rangle/N$ & tricritical point &
  App.~\labelcref{sub:tax_app} \\
  higher-order $SIRS$ & 3 & 4 & $1 - \langle n_S \rangle/N$ & tricritical point &
  App.~\labelcref{sec:app:reducedoad} \\  
  \bottomrule
\end{NiceTabular}
\caption{%
  \textbf{Overview of the models we study with our formalism, in order of appearance.}
  For each model, we list the number of compartments, the number of elementary reactions, the compartment related to the order parameter (with its average taken at the stationary limit, $t\rightarrow \infty$), and the type or order of the phase transitions that we identified with our approach.
  \textsuperscript{a}The predator-prey model is a spatial variant of the Lotka-Volterra model and consists of prey species $A$, its predator $B$, and empty patches $E$.
  When prey goes extinct ($n_A=0$), predators will follow, hence prey population serves well as the order parameter.
  \textsuperscript{b}The generalized Lotka-Volterra is a general model with $M$ compartments (species) and $M\times (M+1)$ reactions and, without loss of generality, any of the compartments can be chosen to define the order parameter --- here we choose $n_1$.}
\label{tab:modeloverview}
\end{table*}

%%%
\section{Results and discussion}\label{sec:results}
Having put forward the theoretical foundations of our formalism, let us now demonstrate its power and flexibility by highlighting its ability to describe a broad spectrum of critical phenomena with both theoretical and practical significance across diverse CRNs from multiple fields (\cref{tab:modeloverview}).
In~\cref{subsec:SIS}, we begin by illustrating how to handle continuous phase transitions. 
We then proceed, in~\cref{subsec:modifiedSIS}, to explore the emergence of a tricritical point and a discontinuous transition mediated by higher-order interactions. 
Finally, in~\cref{sec:spatiallotkavolterra}, we address a case involving degenerated parametrizations. 
Further details about the phenomenology encountered in these sections are provided in~\cref{sec:app:spatiallotkavolterra,sec:app:reduced,sub:tax_app}.

\subsection{Continuous phase transitions: The \texorpdfstring{$SIS$}{SIS} model}\label{subsec:SIS}%
The well-known $SIS$ model has been addressed in several works using the WKB formalism (e.g.,~\cite{khasin2009extinction}) to investigate its stochastic properties. Here, we employ it to illustrate how to identify second-order phase transitions through the Hamiltonian representation. In the $SIS$ model, a susceptible individual ($S$) can become infected upon the interaction with infected individuals ($I$) with some rate $\gamma/N$, where $N$ is the total number of individuals at $t=0$ and is constant.
We consider the closed system wherein infected individuals can recover with some rate $\rho$ to become susceptible again.
This means that we consider the elementary reactions (the CRN) to be
\begin{subequations}
    \label{eq:sis}
    \begin{align}
        S + I &\xrightarrow{\gamma/N} 2I \\
        I &\xrightarrow{\rho} S
    \end{align}
\end{subequations}
Using the creation and annihilation operators (\cref{eq:operators}) we find the DPH to be (see also~\cref{eq:app:crntodph})
\begin{equation}
    \label{eq:dphsis}
    \mathcal{H} = \frac{\gamma}{N} a_I^\dagger (a_I^\dagger - a_S^\dagger) a_I a_S + 
    \rho (a_S^\dagger - a_I^\dagger) a_I
\end{equation}
Next, we replace the operators by their respective real fields (see~\cref{subsec:wkb}), i.e.,~$a_i^\dagger \Leftrightarrow x_i$ and $a_i \Leftrightarrow q_i$.
This makes the Hamiltonian read
\begin{equation}
    \mathcal{H} = \frac{\gamma}{N} x_I(x_I - x_S) q_I q_S + \rho (x_S - x_I) q_I
\end{equation}
Next, recall that the total population is conserved, i.e., $n_S + n_I = N$, with $n_i = x_i q_i$ the population number of the considered compartments.

%% FIGURE
\begin{figure}[b]
    \centering
    \includegraphics[width=\linewidth]{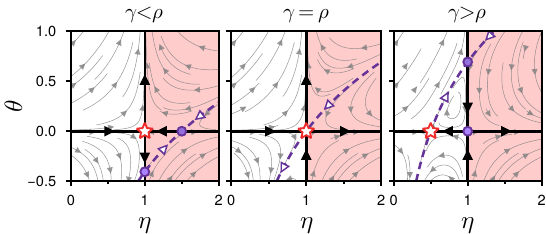}
    \caption{
        \textbf{Second-order phase transition of the $SIS$ model}.
        Phase portrait in the $(\eta,\theta)$-plane of the $SIS$ model with recovery rate $\rho$ and  infection rate $\gamma$.
        (left)~When $\gamma<\rho$, only $\eta=1$ is the global attractor on the mean-field manifold $\theta=0$, as $\eta>1$ is forbidden as it rests in the nonphysical region (see the text and~\cref{fig:intro}).
        (middle)~When $\gamma \rightarrow \rho$, the fixed points on the mean-field manifold converge continuously until they have merged at $\gamma=\rho$, and (right)~when $\gamma>\rho$ the global attractor lies on the intersection with the mean-field manifold at $\eta = \rho/\gamma$.
        As there exists no discontinuous change of the global attractor, this phase transition is of second order.
        For more clarification on the phase plots, see~\cref{fig:intro}.
        The parameters used are $N=1$ and $\rho=1$.
    }
    \label{fig:sis}
\end{figure}

Now we need to freeze the infectious compartment, which means to apply the constraints
\begin{equation}
    x_I = 1, \qquad q_I = N - x_S q_S
\end{equation}
In combination with the Cole-Hopf transformations, we write for $\boldsymbol{x}$ and $\boldsymbol{q}$ the following relations
\begin{subequations}
    \begin{align}
        x_S &= e^\theta, & q_S &= \eta e^{-\theta}, \\
        x_I &= 1,  & q_I &= N - \eta,
    \end{align}
\end{subequations}
where we have suppressed the subscript, i.e.,~$\eta \equiv \eta_S$ and $\theta \equiv \theta_S$.
This gives us the reduced Hamiltonian
\begin{equation}
    \mathcal{H}^\star = (1 - e^\theta)(N-\eta)
    \bigg(\frac{\gamma\eta}{N}e^{-\theta} - \rho\bigg)
\end{equation}
Note again that this Hamiltonian describes a \emph{reduced} system in the $(\eta,\theta)$ plane, as we have effectively frozen the infectious compartment by setting $x_I=1$, and have thereby reduced the dimensions of the problem.
We can extract the zero-energy lines of the reduced Hamiltonian, i.e., the curves at which $\mathcal{H}^\star = 0$.
They occur for $\eta=N$, $\theta=0$, and 
\begin{equation}
    \theta = \log\left( \frac{\eta\gamma}{\rho N} \right)
\end{equation}
We call the zero-energy lines for which $\eta=N$ and $\theta=0$ the \emph{trivial zero-energy lines}, as they are easily obtained by taking a glance at the Hamiltonian and correspond to a trivial fixed point (typically fixation, at $\eta=N$, or extinction, at $\eta=0$) and the mean-field approximation (at $\theta=0$, see~\cref{eq:meanfielddynamics}).
On the trivial zero-energy line at $\theta=0$, we find two fixed points, given by $F_1 = (N,0)$ and $F_2 = (\rho N / \gamma, 0)$.
The third fixed point is at the intersection of the curve $\theta(\eta)$, which we call a \textit{nontrivial} zero-energy line, with $\eta=N$ and, thus, we obtain $F_3 = (N,\log(\gamma/\rho))$.

%% FIGURE
\begin{figure*}[t]
    \centering
    \includegraphics[width=\linewidth]{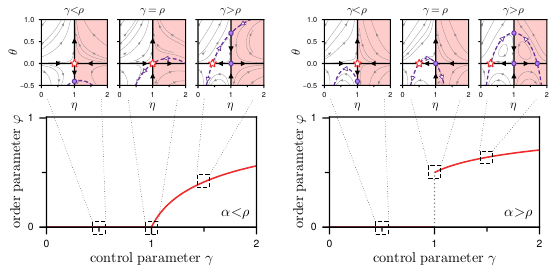}%
    \caption{%
        \textbf{Analysis of the modified $SIS$ model displays first-order phase transition}.
        The order parameter $\varphi = 1 - \langle n_S \rangle$ versus the control parameter $\gamma$ for (left)~$\alpha<\rho$, and (right)~$\alpha>\rho$, on the mean-field manifold $\theta=0$ ($x_I=1$, see the text).
        For some values of $\gamma$, we have drawn the phase portrait in the $(\eta,\theta)$ plane. 
        On the left, one can observe that the global attractor (red star) changes continuously as the nontrivial zero-energy line (dashed purple) crosses the trivial one, and thus the system admits a second-order phase transition at $\gamma=\rho$ when $\alpha<\rho$.
        Instead, on the right, the global attractor changes abruptly at $\gamma=\rho$ as the leftmost attractor becomes the new global attractor, which thus indicates a first-order phase transition.
        Note that these phase portraits can be observed \emph{without} solving or simulating the dynamics themselves.
        For more clarification on the phase plot, see~\cref{fig:intro}.
        The parameters used are $N=1$, $\rho=1$, $\alpha=0.5$ (left), and $\alpha=2$ (right).
    }%
    \label{fig:modifiedsis}
\end{figure*}

To gain information on the critical behavior, we introduce the order parameter $\varphi = 1- \langle n_S \rangle/N$ and the control parameter $\gamma$, and investigate the phase portrait in the $(\eta,\theta)$ plane, which is depicted in~\cref{fig:sis}.
From the phase portrait, we can see that the global attractor changes smoothly as we vary the infection rate $\gamma$, thus indicating that the phase transition is of second order, as expected. 
Note that, in~\cref{fig:sis}, we have imposed the normalization $N=1$.
This is equivalent to saying that we consider instead the population density $n/N$, which, for $N\gg 1$, can be approximated as a continuous variable whose dynamics is described by the same mean-field equations as $n$.

This simple model illustrates how one should go about applying the dimension reduction scheme, and how to analyze the phase portraits in order to reason about the nature of the phase transition.

\subsection{Emergence of discontinuous transitions: A modified \texorpdfstring{$SIS$}{SIS} model}\label{subsec:modifiedSIS}%
At odds with second-order phase transitions, there exist scenarios where abrupt changes occur in the structure of the critical points of phase portraits upon the variation of some parameters. As we see next, this behavior hints to first-order, discontinuous phase transitions.

To illustrate this, we modify the $SIS$ model by including an additional reaction that helps spread the infection. In addition to the reactions in~\cref{eq:sis}, we also consider the reaction
\begin{equation}
    S + 2I \xrightarrow{\alpha/N^2} 3I
\end{equation}
Such a reaction facilitates the spread of infections by means of a higher-order interaction.
Do note that in this case, whether a fixed point of the dynamics is globally attracting or not may depend on the initial conditions (see~\cref{sec:app:initialconditions}).
However, as we assume $n_S \approx N$ at $t=0$, the fixed points that we investigate here always correctly reflect the dynamics.

When including the additional reaction, the DPH for this system now reads
\begin{equation}
    \mathcal{H} = \mathcal{H}_{SIS} +
    \frac{\alpha}{N^2} (a^\dagger_I)^2 (a_I^\dagger - a_S^\dagger) a_I^2 a_S,
\end{equation}
with $\mathcal{H}_{SIS}$ the Hamiltonian of the $SIS$ model, as in~\cref{eq:dphsis}.
As before, we freeze the infectious compartment, replace the creation and annihilation operators by their respective fields, apply the Cole-Hopf transformation, and obtain the reduced Hamiltonian
\begin{equation}
    \mathcal{H}^\star = (1-e^{-\theta})(N-\eta)      
    \left(
        \zeta \eta e^{-\theta} - \rho
    \right)
\end{equation}
where we have defined
\begin{equation}
    \zeta = \frac{N(\alpha+\gamma) - \alpha \eta}{N^2}
\end{equation}
As in the $SIS$ model, the trivial zero-energy lines are obtained for $\theta=0$ and $\eta=N$, but now the nontrivial zero-energy line is given by the curve
\begin{equation}
    \theta = \log \left( 
    %\frac{\eta \left[ N(\alpha+\gamma) - \alpha\eta \right]}{\rho N^2}
    \frac{\zeta \eta}{\rho}
    \right),
\end{equation}
which crosses the trivial zero-energy line at $\theta=0$ at values of $\eta$ that depend on the system parameters~(\cref{fig:modifiedsis}).
At the critical point $\gamma=\rho$, we find that, after normalization ($N=1$) these values are
\begin{equation}
    \eta = 1, \qquad \eta=\rho/\alpha,
\end{equation}
and, hence, one can see that if $\alpha > \rho$ a new fixed point in the physical region emerges, thereby indicating a first-order transition.
On closer inspection of the order parameter, this phenomenon is indeed confirmed~(\cref{fig:modifiedsis}).
As such, this model illustrates how one can expect the phase portraits to resemble when the system at hand displays discontinuous phase transitions.

The last two sections have been presented for illustrative purposes. 
In~\cref{sub:tax_app}, we provide a detailed explanation of how our dimension reduction method can be applied to a tax evasion model featuring a more complex CRN~\cite{tax}, which exhibits both first- and second-order transitions.

\subsection{Overcoming parametrization degeneracy in a simple predator-prey model}
\label{sec:spatiallotkavolterra}% 
As anticipated in~\cref{sec:theoretical_reduction}, finding the solutions of~\cref{eq:integraliprimi} after imposing the constraints may lead to different parametrizations.
In this section, we introduce an example in which two possible parametrizations arise and show how to use physical reasoning to find the correct one that reproduces the actual critical behavior in a specific regime.

For this endeavor, we use an ecological metapopulation model. Let us assume that, in this model, there is a limited number of patches $N$.
We consider two distinct species that inhabit these patches, one of the prey species $A$, and one of its predator $B$.
We shall furthermore introduce a factor $D<N$ that represents the amount of destroyed, uninhabitable patches (cf. Ref.~\cite{sole2024nonequilibrium}, i.e.,~the total number of available patches equals $N-D$).
Note that, while we focus mainly on the parametrization and the characterization of phase transitions, this model is additionally of ecological interest, as its analysis will allow one to reason about the effects of habitat loss on species maintenance.
This aside, we assume that prey species colonize empty patches $E$ with rate $c$ and they are predated upon with rate $k$.
For simplicity, we consider that both the predator and the prey die with rate $\varepsilon$.
This system is, thus, described by the elementary reactions
\begin{subequations}\label{eq:spatiallotkavolterra}
  \begin{align}
    A + E &\xrightarrow{c/N} 2A \\
    A + B &\xrightarrow{k/N} 2B \\
    A &\xrightarrow{\varepsilon} E \\
    B &\xrightarrow{\varepsilon} E
  \end{align}
\end{subequations}
Note that these reactions can be interpreted as reactions that describe a spatially explicit Lotka-Volterra model, where $N-D$ represents the carrying capacity of the prey species.
However, as we work under a well-mixed assumption, these reactions instead describe a simplified, space-agnostic version of the stochastic lattice Lotka-Volterra model (see, e.g., Refs.~\cite{mobilia2007phase, dobramysl2018stochastic}).

\begin{figure*}[t]
    \centering
    \includegraphics[width=0.95\linewidth]{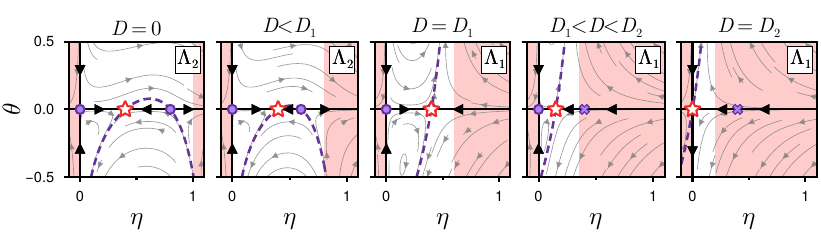}%
    \caption{%
        \textbf{Phase portraits of the Lotka-Volterra model display parametrization degeneracy.}
        Phase portraits in the $(\eta,\theta)$ plane of the Lotka-Volterra model (\cref{eq:spatiallotkavolterra}) for different amounts of habitat loss and different parametrizations $\Lambda_1$ and $\Lambda_2$ (see the text).
        Zero-energy lines defined by the reduced Hamiltonians that are obtained when using a specific parametrization (i.e., $\mathcal{H}_{1,2}^\star = \mathcal{H} \circ \Lambda_{1,2})$.
        (left)~When $D < D_1$, both species persist, and we find that all fixed points are physically meaningful and that they are all correctly identified by $\Lambda_2$. On the other hand, the parametrization $\Lambda_1$  fails to capture one of the fixed points ($F_2$, right purple point) and hence cannot be used to characterize the phase transition in this regime (see the text and~\cref{sec:app:spatiallotkavolterra} for more details).    
        (right) When $D_1 \leq D < D_2$, the predator species goes extinct and only the prey species remains.
        Despite the parametrization $\Lambda_2$ describing the correct critical behavior, it does not include the absorbing conditions of the equilibria, therefore it displays a nonphysical fixed point (purple cross) and $\Lambda_1$ must be used instead.
        Finally, when $D\geq D_2$, both species go extinct and $\eta=0$ remains the only physical attractor.
        Note that the nonphysical domain (red) changes depending on the control parameter $D$ as fewer patches become available.
        For more clarification on the phase plot, see~\cref{fig:intro}.
        Parameters are $c=1$, $k=0.5$, and $\varepsilon=0.2$, which define $D_1=0.4$ and $D_2=0.8$, and from left to right we have $D=0, 0.15, 0.4, 0.65$ and $0.8$.
    }%
    \label{fig:spatiallotkavolterraphaseplot}
\end{figure*}

Using these reactions, we construct the Doi-Peliti Hamiltonian, which reads
\begin{align}
  \label{eq:levinhamiltonian}
  \begin{split}
    \mathcal{H} &= \frac{c}{N} a_A^\dagger(a_A^\dagger - a_E^\dagger) a_A a_E 
                    + \varepsilon (a_E^\dagger - a_A^\dagger) a_A \\
                &\quad + \frac{k}{N} a_B^\dagger(a_B^\dagger - a_A^\dagger) a_B a_A 
                + \varepsilon (a_E^\dagger - a_B^\dagger) a_B.
  \end{split}
\end{align}
Following the steps of our dimension reduction scheme, we consider the control parameter $D$ and the order parameter $\varphi = 1 - \langle n_A \rangle / N$.
We freeze all other compartments by letting $x_{B,E} = 1$, and note that one of the constraints on the occupation number is given by
\begin{equation}\label{eq:totalpop}
  N-D = n_A + n_B + n_E.
\end{equation}
 This constraint alone is not sufficient to reduce the system to a one-dimensional model, and,  thus, we have to include the mean-field dynamics obtained by solving the Hamilton equations.
That is to say, we look for stationary trajectories on the mean-field manifold $\boldsymbol{x}=1$, which are those that abide
\begin{equation}
  \label{eq:4}
  \partial_t \langle n_j \rangle = \left[ \frac{\partial\mathcal{H}}{\partial x_j} \right]_{\boldsymbol{x}=1}
\end{equation}
After applying the substitution $q_j = \langle n_j \rangle$ (\cref{eq:meanfielddynamics}), the mean-field dynamics read
\begin{subequations}
  \begin{align}
    \partial_t \langle n_A \rangle &= \frac{c}{N} \langle n_A \rangle \langle n_E \rangle - \frac{k}{N}\langle n_A \rangle \langle n_B \rangle 
    - \varepsilon \langle n_A \rangle \label{eq:nA} \\
    \partial_t \langle n_B \rangle &= \frac{k}{N} \langle n_A \rangle \langle n_B \rangle - \varepsilon \langle n_B \rangle \label{eq:nB} \\
    \partial_t \langle n_E \rangle &= - \big( 
        \partial_t \langle n_A \rangle +
        \partial_t \langle n_B \rangle 
    \big) \label{eq:nE}
  \end{align}
\end{subequations}
By solving these equations and including the population conservation of~\cref{eq:totalpop}, we obtain the system's fixed points $F=(n_A,n_B,n_E)$ as
\begin{subequations}
    \label{eq:slvfixedpoints}
    \begin{align}
        F_0&=( 0,0,N-D) \\
        F_1&=\left(N-D-\frac{\varepsilon N}{c}, 0,\frac{\varepsilon N}{c}\right) \\
        F_2&=\left(\frac{\varepsilon N}{k},\frac{c(N-D)}{c+k}-\frac{\varepsilon N}{k}, \frac{k(N-D)}{c+k}\right)
    \end{align}
\end{subequations}

We now apply our framework to analyze the critical behavior. 
As discussed above, the image of the parametrization must include all the fixed points of the complete system, which, in this case, depend on the order parameter $D$ (\cref{sec:app:spatiallotkavolterra}).
We can see that, solving the system given by the stationary equation for 
the predators $\partial_t\langle n_B\rangle=0$ (\cref{eq:nB}) and using~\cref{eq:totalpop},
the first parametrization obtained after normalization ($N=1$) reads
\begin{align}
    \begin{split}
            \Lambda_1 = \lbrace & x_A,\; x_B=1,\; x_E=1, \\
              & q_A,\; q_B=0,\; q_E= 1-D - q_A x_A \rbrace
    \end{split}
\end{align}
The corresponding reduced Hamiltonian $\mathcal{H}_1^\star = \mathcal{H} \circ \Lambda_1$ is easily obtained and reads
\begin{equation}
    \mathcal{H}_1^{\star} = \eta (1 - e^\theta) (\varepsilon e^{-\theta} + c \chi),
\end{equation}
where we have used $\chi = c(\eta + D - 1)$.
However, note that this parametrization captures only the states where predators are extinct as $q_B=0$.
This is a consequence of~\cref{eq:nB} having a null kernel, which leads to $q_B=0$ when using $(x_A,q_A)$ as variables.

When using
the stationary equation for the empty patches $E$ instead (\cref{eq:nE}), we find another parametrization $\Lambda_2$ that \emph{does} include all the fixed points (see also~\cref{fig:app:spatiallotkavolterraparam2}), i.e.,
\begin{align}
    \begin{split}
        \Lambda_2 = \lbrace 
            &x_A,\; x_B = 1, \; x_E=1,\; \\ &q_A, \\
            &q_B = \frac{c q_A (1 - D - x_A q_A) - \varepsilon q_A}{c q_A + \varepsilon}, \\
            &q_E = \frac{1 - D - x_A q_A - q_A}{c q_A + \varepsilon} \rbrace
    \end{split} 
\end{align}
We can then derive the corresponding reduced Hamiltonian $\mathcal{H}_2^{\star}=\mathcal{H}\circ \Lambda_2$, which reads
\begin{equation}
    \label{eq:HredspatialLV}
    \mathcal{H}^\star_2 = \eta (1 - e^\theta)\mathcal{H}_{\textrm{nt}},
\end{equation}
with 
\begin{equation}
    \mathcal{H}_{\textrm{nt}} = 
    \frac{\left(
        \chi \varepsilon e^\theta+ 
        \varepsilon (\varepsilon - c\eta) - 
        \eta e^{-\theta} \left[
        \varepsilon k - \varepsilon c + \chi k
        \right]
    \right)}{{c\eta + \varepsilon e^\theta}}
\end{equation}
the part of the Hamiltonian that defines the nontrivial zero-energy line and $\chi$ as defined above.
We again emphasize that this reduced Hamiltonian captures all fixed points through its nonzero-energy lines (\cref{fig:app:spatiallotkavolterraparam2}), while the previous parametrization did not.

Next, using the fixed points (\cref{eq:slvfixedpoints}) we can identify three distinct regions in the phase space that depend on the value of the control parameter $D$.
These regions are separated at $D_1$ and $D_2$, where the first is obtained by checking for what value of $D$ one finds $F_2=F_1$ (predator extinction, i.e.~$n_B=0$), and the second is obtained by checking for what value of $D$ one gets $F_1=F_0$ (full extinction, i.e.~$n_{A,B}=0$).
In doing so, we find two critical values for the order parameter (see also~\cref{sec:app:spatiallotkavolterra} for more details)
\begin{equation}
  D_1 = 1 - \frac{\varepsilon}{k} - \frac{\varepsilon}{c}, \qquad 
  D_2 = 1 - \frac{\varepsilon}{c}
\end{equation}

By studying the phase portrait (\cref{fig:spatiallotkavolterraphaseplot,fig:app:spatiallotkavolterraparam2}), we note that, as the global fixed point varies smoothly when increasing $D$, and since no other fixed point appears at $D_1$ nor at $D_2$, both phase transitions are second-order transitions. 
Still, when $D_1<D<D_2$, we see that the fixed point corresponding to $F_2$ remains (\cref{fig:app:spatiallotkavolterraparam2}), yet in this regime the prey population $n_B$ becomes negative.
Thus, although $\Lambda_2$ correctly identifies all fixed points of the dynamics --- which, importantly, allow us to characterize the phase transition --- its accompanying phase portrait includes a nonphysical fixed point.
This phenomenon arises from the fact that the absorbing boundary condition at predator extinction (i.e., when $F_1=F_2$ for $D=D_2$) is not properly taken into account.
To recover the actual dynamics in this parameter regime, we must instead choose $\Lambda_1$ as the parametrization, as we have done in~\cref{fig:spatiallotkavolterraphaseplot}.

Before closing this section, we briefly expand on the ecological intuition behind these results.
We have seen that, for some levels of destroyed habitat, both the predator and its prey can coexist.
When the available habitat decreases, the predator eventually goes extinct, and only the prey remains.
Further destruction of suitable habitat eventually leads to the extinction of the prey as well. These results are seen when integrating the (mean-field) dynamics as well (see~\cref{sec:app:spatiallotkavolterra}). This model, thus, highlights that fragmentation is generally detrimental toward coexistence and system stability, and the fact that these extinctions occur through phase transitions at finite values of $D$ illustrates the fragility of ecosystems exposed to these kinds of phenomena~\cite{kefi2024self}. 
We consider a more thorough investigation on the ecological ramifications of this model out of the scope of this manuscript, and we refer those interested to related studies on phase transitions in ecological systems to Refs.~\cite{bascompte1996habitat, scheffer2001catastrophic, suding2009threshold, sole2024nonequilibrium}.

Concretely, let us summarize parametrization degeneracy and how it may be resolved.
When studying a CRN for which multiple parametrizations arise when reducing its dimensions, it is important to know that not all parametrizations are equal and that they may not correctly represent the true dynamics, or they may not identify all fixed points of the system. 
In order to characterize the phase transition, we must choose the parametrization that allows us to visualize \emph{all} fixed points, being certain that the possible absorbing states are being taken into account when these become relevant.

In fact, absorbing states drastically change the system.
For example, fluctuations in prey abundance originating from predation simply cease to exist because there are no predators.
Hence, a different system is to be studied instead, which often requires a change in parametrization and, subsequently, the choice of a different reduced Hamiltonian, which leads to a different phase portrait.
We close this section by noting that, even if we successfully resolve the degeneracy and characterize the phase transition for this predator-prey model, a significant open challenge remains, namely, addressing the emergence of degeneracy in its full generality and developing a systematic approach to handling absorbing boundary conditions.

\subsection{Effective dynamics in the absence of spontaneous transitions}%
\label{subsec:fluctuations}
Finally, we highlight that the effective one-dimensional dynamics can provide information beyond the mean-field description of the dynamics, as we argued in~\cref{sec:reduced}.
Recall that we approximate the true (high-dimensional) stochastic dynamics, as described by the chemical Langevin equation (\cref{eq:chemicallangevin}), using a Langevin equation wherein both the drift and noise strength are captured by the reduced Hamiltonian (\cref{eq:efflangevin}).
In addition, we noted that spontaneous transitions manifest themselves as off-diagonal entries in the correlation matrix of the original (nonreduced) problem, and that our dimension reduction scheme fails to capture fluctuations arising from these kind of transitions (see also~\cref{sec:reduced,sec:app:reduced}).

%% FIGURE
\begin{figure}[b]
  \centering
  \includegraphics[width=.95\columnwidth]{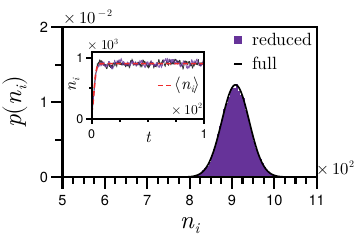}
  \caption{%
    \textbf{Reduced dynamics capture stationary fluctuations.}
    Comparison between typical realizations of the full dynamics and the reduced dynamics for the generalized Lotka-Volterra model with $M=24$.
    The inset shows typical trajectories of the full dynamics (black line), reduced dynamics (purple line), and the mean-field dynamics (red line).
    Note that the reduced dynamics captures both the mean-field dynamics and the fluctuations about a feasible steady state, as the distributions of fluctuations of both systems are nearly indistinguishable.
    See~\cref{sec:app:reducedglv} for additional details and parameters.
  }
  \label{fig:glvhisttraj}%
\end{figure}

However, CRNs \emph{without} spontaneous transitions are just as plentiful as those with, and our reduced Langevin equation in fact fully captures fluctuations around the fixed points, even for systems with a high number of distinct compartments.
To illustrate this, we consider a well-known and well-studied ecological model, the generalized Lotka-Volterra model, and focus on the fluctuations about a steady state. 
Briefly, the generalized Lotka-Volterra model pertains an ecological system of $M$ interacting species with abundances $N_i$, and is described by the following set of $M\times (M+1)$ elementary reactions~\cite{giuli2022dynamical}:
\begin{subequations}
  \begin{align}
    N_i &\xrightarrow{r_i} 2 N_i \\
    N_i + N_i &\xrightarrow{r_i / \kappa_i} N_i \\
    N_i + N_j &\xrightarrow{\alpha_{ij}/\kappa_i} N_j
  \end{align}
\end{subequations}
where $r_i$ the growth rate, $\kappa_i$ the carrying capacity, and $\alpha_{ij}$ the coupling coefficient that defines the (negative) effect of species $j$ on species $i$ (note that $\alpha_{ii}=0$).
Note the absence of spontaneous transitions as species growth and competition between distinct species always depends on the focal species involved.

One can follow the steps of our dimension reduction scheme to obtain the effective, or reduced, Langevin equation (\cref{sec:app:reducedglv}), and comparison of numerical integration of the full system and the reduced one reveals essentially identical fluctuation statistics (\cref{fig:glvhisttraj}).
That is to say, in the absence of spontaneous transitions, the effective one-dimensional dynamics accurately captures the original systems' fluctuations, illustrating that our dimension reduction scheme captures stochastic phenomena well beyond a mean-field description, even for complex systems with many compartments.
%

%%%
\section{Conclusions}\label{sec:conclusions}%
In this work, we have presented a unifying framework that shows how to systematically apply the Doi-Peliti formalism to a broad class of compartmental models. 
Several studies have proven how path integral approaches can offer advantages when describing stochastic systems, characterizing phase transitions, and giving information beyond the deterministic limit. 
Yet, dealing with systems with many compartments is generally complicated, especially regarding the visualization of the geometry of phase portraits to identify the nature of the phase transitions. 
Here, we have derived a novel strategy that allows us to unify these descriptions thanks to the introduction of a dimension reduction scheme. 
In practice, our method provides information about the nature of the phase transitions, and we can analyze both qualitatively and quantitatively the stochastic dynamics. 
We do so through a transparent physical interpretation: Since the observable of interest, the order parameter, corresponds to a single compartment, one can freeze all other degrees of freedom and study only the compartment corresponding to the observable. 

We have furthermore illustrated the usefulness of this methodology by applying it to several relevant multicompartmental stochastic processes of increasing complexity and of different nature, including epidemiological, ecological and sociotechnical ones. 
Starting from the simplest case of two-compartment systems, we explored the role of three-body interactions in epidemic-inspired models, which have recently been reported to significantly impact the behavior of many empirical systems~\cite{battiston2021physics}.
This opens the door to systematically analyze systems whose interactions are beyond pairwise. 
For systems with more than two compartments, our dimension reduction proposal offers very good approximations for the stationary probability density function of the order parameter and successfully captures their critical properties (\cref{sec:app:reduced}).

To derive our dimension reduction scheme, we have used some approximations that do not impact the predictions of the critical behavior, yet they do have an effect on the transient dynamics. 
Indeed, many methods that reduce the dimensions of a system aim to describe not only the stationary limit but also to provide information about the dynamics, which is, however, often at the expense of needing more assumptions. 
Hence, there is a compromise between the amount of information one can retrieve and the generality of the model one can actually study. 
Our method positions itself in a place in which it can be generally applied to a range of systems that is wider than what traditional methods offer, such as those based on timescale separation~\citep{slowchem, slow}, but misses the possibility of an accurate prediction of the transient phase.
Offering insights on the transient dynamics from the point of view developed here remains a topic for future research. 
Additionally, our method is most useful in scenarios in which only one degree of freedom is effectively operating in the long-time behavior, so it is not yet clear to what degree we can freely reduce the dimensionality in cases with coupled degrees of freedom.

Another aspect that we believe is worth further consideration is related to the parametrization $\Lambda$ and the reduced Hamiltonian $\mathcal{H}^{\star}$.
More specifically, the difficulties are associated with the definition of the validity domain and the potential degeneration of $\Lambda$. 
An open question that remains that we foresee to have a great impact due to the wealth of potential applications, is the generalization of this dimension reduction scheme to structured, networked populations that interact beyond the all-to-all limit~\cite{barrat2008dynamical}.

In conclusion, we have exploited the combination of the path integral representation of stochastic processes with a dimension reduction scheme to better understand the critical properties and the stationary behavior of coupled compartmental stochastic processes. We hope that our formalism will provide new perspectives and physical insights into, and beyond, the characterization of phase transitions and fluctuations of multicompartment complex systems.

\section*{\label{sec:Acks}Acknowledgments}

We thank the anonymous referees for their constructive feedback.
G.V.V. and J.N. acknowledge financial support from the Human Frontier Science Program Organization (HFSP Ref. RGY0064/2022).
T.S. acknowledges financial support from MUR funding within the FIS (DD n. 1219 31-07-2023) Project no. FIS00000158.
O.A. acknowledges financial support from the Spanish Grant No. PID2021-128005NB-C22, funded by MCIN/AEI/10.13039/501100011033, ``ERDF A way of making Europe'', from Generalitat de Catalunya (2021SGR00856), the Spanish Ministry of Universities through the Recovery, Transformation, and Resilience Plan funded by the European Union (Next Generation EU), and the University of the Balearic Islands.
M.D.D. acknowledges partial financial support from MUR within the PRIN 2022 PNRR (DD n. 1214 31-07-2023), Project no. P2022A889F. 

%\nocite{*}
\bibliography{biblo}

\clearpage
\beginsupplement
\appendix

\section{Details for the derivation of the Doi-Peliti Hamiltonian}\label{sec:DPHforCRN_app}
Here, we demonstrate how one constructs the Doi-Peliti Hamiltonian (DPH) for chemical reaction networks. 
For most chemical reaction networks, including the ones within the scope of our work, we can identify five fundamental processes (elementary reactions).
Each of these processes are automatically mapped to a combination of operators, according to the definitions in~\cref{eq:operators}, which appear in the DPH. 
One can show that they correspond to
\begin{subequations}
    \label{eq:app:crntodph}
    \begin{align}
        &\mathrm{\textbf{elementary\;reaction}} \qquad &\mathrm{\textbf{DPH}} \nonumber \\
        &\emptyset \rightarrow A \qquad & a_A^\dagger - 1 \label{eq:birthprocess} \\
        &A \rightarrow \emptyset & (1 - a_A^\dagger) a_A \\
        &A \rightarrow 2A \qquad & a_A^\dagger(a_A^\dagger - 1) a_A \\
        &A \rightarrow B \qquad & (a_B^\dagger - a_A^\dagger) a_A \label{eq:transformprocess} \\
        &A+B \rightarrow A+C \qquad & a_A^{\dagger}(a_{C}^{\dagger}-a_B^{\dagger})a_{A}a_{B}
    \end{align}
\end{subequations}
where we have considered the rates to be implicit --- that is to say, the elementary reactions occur with some rate $\lambda$, and these reactions are considered to be mass-action reactions.

To derive the DPH for a chemical reaction network with multiple transitions, one starts by explicitly writing the master equation (\cref{eq:ME}) that corresponds to the elementary reaction(s), as we now illustrate below.

\subsection{The Doi-Peliti Hamiltonian for birth processes}
Let us focus on the birth process (\cref{eq:birthprocess}). 
We let $n$ denote the amount of $A$, such that one obtains
\begin{equation}
    \partial_t p(n) = \lambda \left[ p(n-1) - p(n+1) \right],
\end{equation}
where the first term corresponds to the probability of arriving in the state $n$ and the second term the probability of leaving $n$.
Then, 
\begin{equation}
    \partial_t \ket{G} = \sum_n \lambda \left[ p(n-1) - p(n) \right] \ket{n},
\end{equation}
and using the definition in~\cref{eq:operators}, we obtain 
\begin{subequations}
\begin{align}
    \partial_t \ket{G} &= \mathcal{H} \ket{G} \\ &=
    \sum_n \lambda a^\dagger_A p(n-1) \ket{n-1}
    - \lambda p(n) \ket{n} \\
    &= \lambda(a_A^\dagger -1) \ket{G}
\end{align}
\end{subequations}
where we have used the definition of the generating function $\ket{G} = \sum_n p(n) \ket{n}$.
Thus, the DPH readily follows and reads $\mathcal{H} = \lambda(a^\dagger_A - 1)$.

\subsection{The Doi-Peliti Hamiltonian for a transformation process}
Reactions involving multiple compartments complicate the structure of the master equation, which is consequently reflected in the DPH. 
As an example, we consider the reactions in~\cref{eq:transformprocess}, whose master equation reads
\begin{equation}
    \partial_t p(\boldsymbol{n}) = \lambda (n_A + 1) p(\boldsymbol{n} + \mathbb{I}_A - \mathbb{I}_B) -
    \lambda n_A p(\boldsymbol{n}).
\end{equation}
By letting $\boldsymbol{n}^\prime = \boldsymbol{n} + \mathbb{I}_A - \mathbb{I}_B$, we write the time derivative of the generating functional as
\begin{align}
    \partial_t \ket{G} = 
    \sum_{\boldsymbol{n}} \lambda (n_A + 1) p(\boldsymbol{n}^\prime) \ket{\boldsymbol{n}} -
    \lambda n_A p(\boldsymbol{n}) \ket{\boldsymbol{n}}.
\end{align}
Again, by applying the definitions of the creation and annihilation operators one finds
\begin{align}
    (n_A + 1) \ket{\boldsymbol{n}} = a_B^\dagger a_A \ket{\boldsymbol{n}^\prime}, \quad
    n_A \ket{\boldsymbol{n}} = a^\dagger_A a_A \ket{\boldsymbol{n}},
\end{align}
which can be used to rewrite
\begin{equation}
    \partial_t \ket{G} = \lambda (a_B^\dagger a_A - a_A^\dagger a_A) \ket{G}.
\end{equation}
Thus, the DPH follows as $\mathcal{H} = \lambda (a_B^\dagger - a_A^\dagger) a_A$, as indicated in~\cref{eq:transformprocess}.
For other reactions, the same steps will yield the appropriate Hamiltonian.

\subsection{The Doi-Peliti Hamiltonian for more general reactions}\label{sec:DPHforCRN_app_subsecKM}
The above derivations can be generalized to systems with several compartments and more complicated interactions, such as those introduced in~\cref{subsec:modifiedSIS}.
We next show how to proceed in more general cases.

Let us first consider a model with two compartments, $A$ and $B$.
The Kramers-Moyal expansion of the master equation reads
\begin{align}
    \partial_ tp(\bm n,t)=\sum_{\substack{m_A,m_B\\ |\bm{m}|>0}}\frac{ (-1)^ {|\bm m|} }{m_A!m_B!}\partial^{m_A}_{ n_A}\partial^{m_B}_{ n_B}W^{m_A,m_B}(\bm n) p(\bm n,t),
\end{align}
 where  $|\bm m|=m_A+m_B$, and 
\begin{equation}
    W^{m_A,m_B}(\bm n)=\int \diff{y_A}\diff{y_B}\, \omega(\bm n+\bm{y}_A+\bm{y}_B|\bm n)y_A^{m_A}y_B^{m_B}.
\end{equation}
 Here, we denote $\bm{y}_{A,B}=y_{A,B}\mathbb{I}_{A,B}$, with $\mathbb{I}_{A,B}$ defined as in the main text and $y_{A,B}$ real variables.  When considering a two-body interaction 
\begin{equation*}
    p A+ q B\xrightarrow{\gamma} (p+k) A+(q-l) B,
\end{equation*}
the transition rate can be written as
\begin{align}
    \omega(\bm n+ \bm{y}_A+\bm{y}_B|\bm n)=\gamma n_{A}^{p}n_B^q\delta(y_A-k)\delta(y_B+l).
\end{align}
Hence,
\begin{equation}
    W^{m_A,m_B}(\bm n)=\gamma n_{A}^{p}n_B^q k^{m_A}(-l)^{m_B},
\end{equation}
and,  by replacing $\partial_n=\mathrm{i}\theta$,  we find an expression for $\mathcal{H}^{\dagger}_{KM}$:
\begin{align}\notag
    \mathcal{H}^{\dagger}_{KM}(\bm \eta,\mathrm{i} \bm \theta)&=\sum_{\substack{m_A,m_B\\|\bm m|>0}} \frac{(\mathrm{i} \theta_A k)^{m_A}(-\mathrm{i}\theta_B l)^{m_B}}{m_A!m_B!}\gamma n_{A}^{p}n_B^q\\&=\gamma n_{A}^{p}n_B^q\left(e^{\mathrm{i} \theta_A k-\mathrm{i}\theta_B l}-1\right).\label{eq:Hd}
\end{align}
On the other hand, the DPH for this reaction is
\begin{subequations}
\label{eq:ch}
\begin{align}
    \mathcal{H}&=\gamma(a^{\dagger}_A)^p (a^{\dagger}_B)^{q-l}\left((a^{\dagger}_A)^k-(a_{B}^{\dagger})^l\right)a_{A}^pa_B^q\\
&=\gamma\left((a^{\dagger}_{A})^k(a_{B}^{\dagger})^{-l}-1\right)(a^{\dagger}_{A})^pa_{A}^{p}(a_B^{\dagger})^qa_B^q
\end{align}
\end{subequations}
By applying the Cole-Hopf transformation (Eqs.~\eqref{eq:colehopf}), we substitute $a^{\dagger}_{A,B}=e^{\mathrm{i}\theta_{A,B}}$ and $a_{A,B}=e^{-\mathrm{i}\theta_{A,B}}\eta_{A,B}$ into Eq.~\eqref{eq:ch} obtaining the same expression of the Kramers-Moyal expansion~\cref{eq:Hd}. 

The above arguments can be extended to more complex and general interactions
\begin{equation}
 \sum_{i=1}^M p_i\, A_i \xrightarrow{\gamma}\sum_{i=1}^{M}(p_i+k_i)\, A_i,
\end{equation}
with $M$ representing the number of compartments $A_i$, and $k_i$ can be any integer. In this case we define the vector $\bm k=(k_1,k_2,\ldots,k_M)$ and the transition rate is given by
\begin{equation}
    \omega(\bm n+\bm y|\bm n)=f(\bm n)\prod_{j=1}^M\delta(y_j-k_j),
\end{equation}
where $f$ a suitable function of the occupation numbers. 
Thus
\begin{align}
    W^{m_1,\ldots,m_M}(\bm n)&=\int \diff{\bm y}\left(\prod_{j=1}^{M} y_j^{m_j}\right)\omega(\bm n+\bm y|\bm n)f(\bm n)\nonumber \\
    &=f(\bm n)\prod_{j=1}^M k_j^{m_j},
\end{align}
and the adjoint of the Kramers-Moyal operator is
\begin{align}\notag
    \mathcal{H}^{\dagger}_{KM}(\bm \eta,\mathrm{i} \bm\theta)&=\sum_{\substack{m_1,m_2,\ldots,m_M\\|\bm m|>0}}\prod_{j=1}^M \frac{\left(\mathrm{i} \theta_j k_j\right)^{m_j}}{m_j!}f(\bm \eta)\\
    &=\left(\prod_{j=1}^{M}e^{\mathrm{i} \theta_j k_j}-1\right)f(\bm \eta).\label{eq:HcompletaKM}
\end{align}
In terms of the Doi-Peliti representation, the corresponding DPH is
\begin{equation}\label{eq:DPHCRN}  \mathcal{H}=\gamma\sum_{i=1}^M\left((a^{\dagger}_i)^{k_i}-1\right)f(a^{\dagger}a),
\end{equation}
which can also be written as a function of the Cole-Hopf variables, 
\begin{equation}
    \mathcal{H}(\bm \eta,\mathrm{i} \bm \theta)=\gamma\left( \prod_{j=1}^Me^{\mathrm{i} \theta_j k_j}-1\right)f(\bm \eta)
\end{equation}
This is equal to the KM operator~\eqref{eq:HcompletaKM}.
\cref{eq:DPHCRN} shows that the mean-field manifold $\bm{\theta}=0$ ($\bm{x}=\bm{1}$) is always a zero of $\mathcal{H}$. 
For a generic CRN, we can build the Doi-Peliti Hamiltonian by simply adding the terms corresponding to each transition.

\section{Details for the derivation of the path integral representation}\label{sec:PI_derivation_app}
Here, we show the derivation of the path integral representation for the generating function $\ket{G}$ using the Doi-Peliti formalism~\cite{weber2017master}. Let us take the simplest one-compartment case $\ket{\bm{n}}=\ket{n}$ with the basis~\eqref{eq:basis}. 
We also take $\mathcal{H}$ in its normal ordered form. 
By writing the time dependency explicitly $\ket{G}=\ket{G(t)}_{x}$ and by splitting $t$ in infinitesimal intervals $t=\epsilon K$ with $\epsilon\to 0$ and $K\to \infty$, we have
\begin{subequations}
\begin{align}
    &\ket{G(t)}_x=e^{\epsilon\mathcal{H}(a^{\dagger},a)}\ket{G(t-\epsilon)}_{x} \\ 
    &= \int_{\mathbb{R}} \diff{x_{K-1}}e^{\epsilon\mathcal{H}(x,\partial_x)}\delta(x-x_{K-1})\ket{G(t-\epsilon)}_{x} + \mathcal{O}(\epsilon^2) \\ 
    &=\int_{\mathbb{R}^2}\frac{\diff{x_{K-1}} \diff{q_{K-1}}}{2\pi}\,e^{\epsilon\mathcal{H}(x_K,\mathrm{i} q_{K-1})} \\
    &\qquad\qquad \times e^{-\mathrm{i} q_{K-1}(x_{K-1}-x_K)}\ket{G(t_{K-1})}_{x_{K-1}} + \mathcal{O}(\epsilon^2) \nonumber\\
    &=\int_{\mathbb{R}^{2K}}\prod_{j={0}}^{K-1}\frac{\diff{x_j}\diff{q_j}}{2\pi}\,e^{\epsilon\sum_{j=0}^{K-1}\Big(\mathcal{H}(x_{j+1},\mathrm{i} q_{j}) - \mathrm{i} q_{j}\frac{x_{j}-x_{j+1}}{\epsilon}\Big)} \nonumber\\
    &\qquad\qquad\qquad \qquad\times \ket{G(t_0)}_{x_0}+\mathcal{O}(\epsilon^2)\\
    &=\int_{[t_0}^{t)} \mathcal{D}[x,q]\,e^{-\mathcal{S}[x,\mathrm{i} q]}\ket{G(t_0)}_{x_0},
\end{align}
\end{subequations}
where in the last equation we have taken the continuous limit by introducing the measure
\begin{equation}
\int_{[t_0}^{t)} \mathcal{D}[x,q]=\lim_{K\to\infty}\int _{\mathbb{R}^{2K}}\prod_{i={0}}^{K-1}\frac{\diff{x_i}\diff{q_i}}{2\pi},
\end{equation}
and $\mathcal{S}$ is the action defined by~\cref{eq:action}. With this notation, the upper and lower bounds of the integral indicate that we include the initial condition $(x_0,q_0)$ within the integration while the final state $(x_K,q_K)$ is excluded.
For the multidimensional case, the derivation is trivially the same.

\section{WKB approximation for CRNs}\label{sub:wkb_app}
Here, we follow the reasoning of~\cite{Black} and demonstrate that similar arguments can be applied to extend the WKB to the case of chemical reaction networks~\eqref{eq:CRN} as well. Let us denote by $\omega_{k}$ the transition rate related to the reaction between $k$ reagents and by $N$ the total population in the system. For illustrative purposes, let us consider a simple contact process 
\begin{equation}
    A_j + A_j \xrightarrow{\beta} A_j.
\end{equation}
The transition rate is proportional to the number of elements, $n_j$, times the probability of finding another element in the heterogeneous network, $(n_j-1)/N$. 
Thus, assuming $n_j,N \gg 0$ and $n_j \ll N$, we have $\omega_2(\textit{\textbf{n}}-\mathbb{I}_j|\textit{\textbf{n}})=\beta n_j(n_j-1)/N\propto n_j^2/N$.
Generalizing the same reasoning for a contact process between $k$ elements, the transition rate is $\omega_k\propto n_{j}^{k}/N^{k-1}$. 
Since the Doi-Peliti Hamiltonian is a polynomial function of $x_j$ and $\mathrm{i} q_j$, we conclude that the term related to the highest power of $\mathrm{i} q_j$ inside $\mathcal{H}$ is proportional to $(\mathrm{i} q_h)^kN^{-k+1}\,(=N^{-k+1}\partial^k_{x_j})$. 
Furthermore, with these rates, we can write the DPH and the action (\ref{eq:action}) as extensive quantities with respect to $N$ as done in~\cite{Black}. 
If we define the concentration variables $\eta_j=n_j/N$, then the transition rate reads
\begin{equation}\label{eq:WKBassum}
    \omega_k\approx\beta \eta_{j}^k N.
\end{equation}
This scaling law for the transition rates is the main assumption for applying the WKB approximation.
Using~\cref{eq:gen0} and~\cref{eq:DPH}, we can see that
\begin{equation}
    H_{n,m}=N h_{nm}(\eta)\,,\qquad \mathcal{S}=N s(\eta),
\end{equation}
with $h_{nm}$ and $s$ being functions of the concentrations only. 
Using the ansatz $\ket{G}=\exp(-N s)$, we obtain
\begin{equation}
\frac{\partial_{x_j}^k \exp(- N s)}{N^{k-1}}=N(-\partial_{x_j}  s)^k+O(1)\approx N(-\partial_{x_j}  s)^k.
\end{equation}
Inserting the same ansatz in Eq.~\eqref{eq:newME}, the relation $\mathrm{i} q_j=-\partial_{x_j}\mathcal{S}$ holds for any $j$ and we find the Hamilton-Jacobi equation
\begin{equation}\label{eq:HJeq}
  \partial_t \mathcal{S}=-\mathcal{H}\left(\textit{\textbf{x}},-\partial_{\textit{\textbf{x}}} \, \mathcal{S}\right),
\end{equation}
with $\partial_{\textit{\textbf{x}}} \, \mathcal{S}=(\partial_{x_1}\mathcal{S},\partial_{x_2}\mathcal{S},\ldots)$. 
As such, for CRNs, we can always take the stationary action defined by the solutions of Eqs.~\eqref{eq:Hsist} and $\ket{G}=\exp(-\mathcal{S})$.

%%%
\section{A basis for the stationary paths}\label{sec:basis_app}
Let us now consider the basis for the WKB approximation presented in~\cite{weber2017master}. 
Considering for simplicity the one-dimensional case and denoting by $\tilde x(t)$ and $\tilde q(t)$ the solutions of the Hamilton equations~\eqref{eq:Hsist}, we take the following basis:
\begin{align}
   & \ket{\tilde n}=\big(\Delta x+\tilde x(t)\big)^ne^{-\tilde q(t)(\Delta x+\tilde x(t))}\\ 
   &\bra{\tilde{n}}=\big(\partial_{\Delta x}+\tilde q(t)\big)^n,
\end{align}
where $\Delta x$ represents the fluctuations of $x(t)$ around the stationary trajectories.  Similarly, we denote the fluctuations of $\mathrm{i} q(t)$ by $\mathrm{i} \Delta q$.   The creation and annihilation operators act according to, respectively,
\begin{align}\label{eq:bat1}
    &a\ket{\tilde n}=n\ket{\tilde n-1}=(\partial_x+\tilde q(t))\ket{\tilde n}\\\label{eq:bat2}
    &a^\dagger\ket{\tilde n}=\ket{\tilde n+1}=(x+\tilde{x}(t))\ket{\tilde n}.
\end{align} 
With this basis, in the derivation of the path integral representation we still have to include the contribution resulting from the time dependence of $\ket{\tilde{n}}$. For the sake of simplicity, we write $\tilde x(t)=\tilde x$ and $\tilde q(t)=\tilde q$, and, by taking the time derivative of $\ket{\tilde n}$, we have
\begin{align}\notag
    \partial_t \ket{\tilde n}&=\Big[n \partial_t\tilde x(\Delta x+\tilde x^{n-1})\\
    &\quad \, -(\Delta x+\tilde x)^n(\partial_t\tilde q(\Delta x+\tilde x)+\tilde q\partial_t \tilde x)\Big]e^{-\tilde q(\Delta x-\tilde x)}\notag\\
    &=n \partial_t\tilde x\ket{\tilde n-1}-\left(\partial_t\tilde q(\Delta x+\tilde x)+\tilde q \partial_t \tilde x\right)\ket{\tilde n}\notag\\\notag
    & = \left(\partial_t\tilde x \,a-\partial_t\tilde q\,a^{\dagger}-\tilde q\partial_t \tilde x\right)\ket{\tilde n}\\\label{eq:time_dep2}&=\mathcal{B}\ket{\tilde n},
\end{align}
where $\mathcal{B}$ is the time evolution operator of the basis. Within this representation, the master equation~\cref{eq:ME} becomes
\begin{align}
\partial_{t}\ket{G}&=\left(\mathcal{H}+\mathcal{B}\right)\ket{G}.
\end{align}
Now, by replacing $(x,\,\mathrm{i} q)$ by $(\Delta x,\,\mathrm{i}\Delta q)$, and by substituting $\mathcal{H}$ with $\mathcal{H}+\mathcal{B}$ in the steps presented in Appendix~\ref{sec:PI_derivation_app}, we obtain the path integral representation with this new basis. 
In detail, the action reads
\begin{align}
    -&\mathcal{S}[\tilde x+\Delta x,\tilde q+\mathrm{i} \Delta q]\nonumber\\&=\int_{t_0}^t \diff{\tau}\, \mathrm{i} \Delta q\partial_\tau \Delta x+\mathcal{H}(\tilde x+\Delta x,\tilde q+\mathrm{i} \Delta q) \nonumber \\
    &\qquad\qquad+\mathcal{B}(\tilde x+\Delta x,\tilde q+\mathrm{i} \Delta q)\notag\\
    &=\int_{t_0}^t \diff{\tau} \,\Big[ \mathrm{i}\Delta q\partial_\tau \Delta x+\mathcal{H}(\tilde x,\tilde q)+\Delta x\partial_{\Delta x}\mathcal{H}(\tilde x,\tilde q)\notag\\
    &\,\,\qquad+\mathrm{i} \Delta q\partial_{\mathrm{i} \Delta q}\mathcal{H}(\tilde x,\tilde q)+\Delta\mathcal{H}+\tilde q\partial_\tau \tilde x+\mathrm{i} \Delta q\partial _\tau \tilde x\notag\\
&\qquad\qquad\quad\quad-\Delta x\partial_{\tau}\tilde q-\tilde x\partial_{\tau} \tilde q-\tilde q\partial_{\tau}\tilde x\Big]\notag\\
    &=\int_{t_0}^t \diff{\tau} \,\Big[ \mathrm{i} \Delta q\partial_\tau \Delta x+\mathcal{H}(\tilde x,\tilde q)+\Delta x\partial_\tau \tilde q-\mathrm{i} \Delta q\partial_\tau \tilde x\notag\\
&\qquad\qquad\quad\quad+\Delta \mathcal{H}+\mathrm{i} \Delta q\partial_\tau \tilde x-\Delta x\partial_\tau \tilde q-\tilde x\partial_\tau \tilde q\Big]\notag\\\notag
&=\int_{t_0}^t \diff{\tau}\,\Big[ \mathcal{H}(\tilde x,\tilde q)+\tilde q\partial_{\tau}\tilde x\Big] -\tilde x(t)\tilde q(t)+\tilde x_0\tilde q_0\\&\qquad\quad\qquad\quad+\int_{t_0}^t \diff{\tau}\, \Big[\mathrm{i} \Delta q\partial_\tau \Delta x+\Delta\mathcal{H}\Big]\notag\\
&\notag =-\tilde{\mathcal{S}}-\tilde x(t)\tilde  q(t)+\tilde{x}_0\tilde q_0\\&\qquad\qquad\,\,+\int_{t_0}^t \diff{\tau}\, \Big[\mathrm{i} \Delta q\partial_\tau \Delta x+\Delta\mathcal{H}\Big],\label{eq:fluttuazioni}
\end{align}
where in the second equality we have expanded $\mathcal{H}(\tilde x+\Delta x,\tilde q+\mathrm{i} \Delta q)$ around the stationary trajectories, and $\Delta \mathcal{H}$ includes all the contributions of orders $\mathcal{O}(\Delta x^2,\Delta q^2,\Delta x\Delta q)$. 
Note that~$\tilde{\mathcal{S}}$ satisfies the Hamilton-Jacobi equation~\eqref{eq:HJeq}.

Let us now rewrite the generating function in terms of the basis $\ket{\tilde n}$,
\begin{align}
\ket{G(t)}_{x}=e^{\tilde x(t)\tilde q(t)}\sum_np(n,t)\ket{\tilde n}\Big |_{\Delta x=0}.
\end{align}
Thus, the path integral expression is
\begin{align}
    &\ket{G(t)}_x =\\
    & e^{-\tilde{\mathcal{S}}}\int_{[t_0}^{t)} \mathcal{D}[\Delta x,\Delta q]\, e^{\int_{t_{0}}^t \diff{\tau} \,\big(\mathrm{i} \Delta q\partial_\tau \Delta x+\Delta\mathcal{H}\big)}\ket{n_0}\Big|_{\genfrac{}{}{0pt}{}{\Delta x(t)=0}{\tilde x(t)=x}} \nonumber.
\end{align} 
If we assume a fixed initial condition, then $\ket{n}_0=\ket{\tilde{n}_0}e^{\tilde x_0\tilde q_0}=e^{n_0 \log{\tilde{x}_0}}$. 
Within the assumptions of the WKB approximation --- namely that the rates scale as in~\cref{eq:WKBassum} and $N\gg1$ --- the fluctuating part does not contribute, and we have $\ket{G}=\exp\big({-\tilde{\mathcal{S}}}\big)$, where $x$ and $\mathrm{i} q$ are replaced by the real trajectories $\tilde x$ and $\tilde q$.

\section{Closing the transition diagrams}\label{sec:closure_app}
The conditions to apply the WKB approximation (e.g. Eq.~\eqref{eq:WKBassum}) are not always sufficient for our formalism to be used to investigate the system critical properties, and additional constraints on the compartment interactions are necessary. Here, we present a mathematical trick that allows one to extract information on the critical behavior in some of these cases. 

Let us take a CRN with $M$ compartments $\{A_1,A_2,\ldots,A_M\}$ wherein only spontaneous transitions and contact interactions occur. 
We assume that $A_M$ is connected to other compartments by a single spontaneous transition
$A_{j} \xrightarrow{\gamma} A_{M}$, 
and, thus, $A_M$ is the termination of the model diagram. 
We consider fixed system size $N$ and we choose the order parameter $\varphi$ as in~\cref{eq:rho}. 
Now, we focus on $A_1$ and $A_M$ to understand if, and when, a phase transition occurs. 

\begin{figure}[t]
  \centering
  \begin{tikzpicture}[%
  every node/.style={
    draw,circle,minimum width=3em,
    font=\large,text opacity=1
    } 
  ]
    \node (A1) at (0,0) {$A_1$};\
    \node[draw=none,anchor=east] at (A1.west) {(a)};
    \node[fill=gray,fill opacity=0.4] (Aeff) at (2,0) 
    {$A_{\textrm{eff}}$};
    \node (AM) at (4,0) {$A_M$};
    \draw[->,thick] (A1) to node[draw=none,midway,above] {$\omega_1$} (Aeff);
    \draw[->,thick] (Aeff) to node[draw=none,midway,above] {$\omega_{\textrm{eff}}$} (AM);
    \node (B1) at (0,-1.5) {$A_1$};
    \node[draw=none,anchor=east] at (B1.west) {(b)};
    \node[fill=gray,fill opacity=0.4] (Beff) at (2,-1.5) 
    {$A_{\textrm{eff}}$};
    \node (BM) at (4,-1.5) {$A_M$};
    \draw[->,thick] (B1) to node[draw=none,midway,above] {$\omega_1$} (Beff);
    \draw[->,thick] (Beff) to node[draw=none,midway,above] {$\omega_{\textrm{eff}}$} (BM);
    \draw[->,black,thick] (BM) to[in=-60,out=-120,looseness=0.5] 
    node[draw=none,midway,below,inner sep=0pt] {$\omega_M$} (B1);
  \end{tikzpicture}
  \caption{%
  Diagrammatic representation of (a)~an $SIR$-like diagram for the effective \textit{open} process, and (b)~an $SIRS$-like diagram for the effective \textit{closed} process.
  }
  \label{fig:closurediagram}
\end{figure}
We rewrite the diagram of the system as in~\cref{fig:closurediagram}(a), where for simplicity we have replaced the $M-2$ compartments $A_{2},\ldots,A_{M-1}$ by an effective one $A_{\text{eff}}$. 
The system is active when $A_{\textrm{eff}}$ becomes asymptotically populated, and we can write
\begin{equation}
    \partial_t\langle n_{\textrm{eff}} \rangle 
    = f(\boldsymbol{\omega},A_1,A_{\textrm{eff}}),
\end{equation}
with $\boldsymbol{\omega}$ representing the transition coefficients of the model, and $f$ is a suitable function depending on the system we are considering. However, the mean-field equations include
\begin{equation}
    \partial_{t}\langle n_{M}\rangle = \omega_j \langle n_j\rangle,
\end{equation}
from which, at the stationary limit, $\langle n_j\rangle=0$. 
The presence of null terms like this one leads to a degeneration of the system of Eqs.~\eqref{sist:crit}. Thus, $\text{det}(\textbf{J})=0$ for any fixed point, and we cannot use the strategy presented in the main text to analyze the critical behavior.
To overcome this problem, we introduce an auxiliary transition rate that closes the model diagram, connecting $A_M$ directly to $A_1$ by means of a spontaneous transition $A_M \rightarrow A_1$.
In doing so, we obtain a $SIRS$-like model, as depicted in~\cref{fig:closurediagram}(b).
Following a classical treatment (as in, e.g., \cite{pastor2015epidemic}), one can show that the $SIR$ and $SIRS$ models have the same critical condition (see also~\cref{subsec:SIS}). 

Moving to more general processes, we can assume, without loss of generality, that $n_{1}\approx N$ for $t\approx 0$, and take as order parameter $\varphi=1-\langle n_1\rangle/N$. This is a general assumption in epidemiology when, for example, an outbreak of a new disease is about to occur and the number of susceptible individuals in a population is very large.
Then, the critical behavior is defined by the system of equations
\begin{equation}\label{eq:sisteff1}    
    \partial_t \langle n_{\textrm{eff}}\rangle = 
    f(\boldsymbol{\omega},A_{\textrm{eff}}), 
    \qquad 
    \partial_{t}\langle n_M\rangle  = \varepsilon \langle n_j\rangle,    
\end{equation}
where we consider that $A_{j}\in A_{\textrm{eff}}$ is the only compartment linked to $A_{M}$ by a spontaneous transition with rate $\varepsilon>0$. 
As we show below, these arguments can be generalized for systems with several terminating compartments. 
As for the $SIRS$ model, by adding a refilling reaction with rate $\omega_M$, the system of~\cref{eq:sisteff1} becomes
\begin{subequations}
\begin{align}
    \partial_t \langle n_{A_{\textrm{eff}}}\rangle &=
    f(\boldsymbol{\omega},A_{\textrm{eff}}),
    \\
    \partial_{t} \langle n_{M}\rangle &= \varepsilon \langle n_{j}\rangle - \omega_M \langle n_M\rangle
\end{align}    
\end{subequations}

Since the phase transition occurs when $A_{\text{eff}}$ changes from 0 to a positive quantity or vice versa, the critical condition is independent of $A_M$ and is determined by the first equation. 
Thus, we conclude that the critical condition for such a system is the same as for the corresponding closed diagram. 
We can generalize these arguments to more complex systems by imposing only two conditions, namely,
\begin{enumerate}[noitemsep,topsep=1pt]
    \item each extremity of the open diagram is linked to the main process by a spontaneous transition;
    \item the closed diagram is given by adding spontaneous transitions from the extremity to the initial empty state.
\end{enumerate}
The first condition ensures the inner part of the process is independent of the population numbers of the extremities. For example, with pair interactions, $f(\boldsymbol{\omega})$ would depend on the population of the ends. The second condition guarantees that one obtains the same diagrammatic structure as the $SIRS$ model.

In conclusion, while the critical behavior of some CRNs might initially appear unsuitable to be tackled with our approach due to the problems in identifying the appropriate parametrization, we demonstrate here that this issue can be effectively overcome by introducing auxiliary rates. However, we emphasize that although they accurately capture the critical behavior, they do alter the system dynamics, such as the trajectory toward the steady state.

%%%
\section{Langevin equation from the Doi-Peliti Hamiltonian}\label{sec:lang_appendix}
By means of the relation of~\cref{eq:re}, we can show the link with the Langevin equation of~\cref{eq:efflangevin}. 
While the procedure for deriving Langevin dynamics from $\mathcal{H}_{\textrm{KM}}$ has been discussed in several works (see, e.g. Refs.~\cite{weber2017master,giuli2022dynamical}); for completeness, we sketch the derivation below. 

In the simple one-dimensional case, we can develop a path integral representation of the probability distribution starting from the Kramers-Moyal expansion of~\cref{eq:KM}.
Following the same steps reported in~\cref{sec:PI_derivation_app}, we write
\begin{equation}
    p(n,t)=\int_{[0}^{t)}\mathcal{D}[\eta,\theta]\,e^{-\mathcal{S}^{\dagger}_{\textrm{KM}}[\eta,\mathrm{i} \theta]}\delta(n-\eta(t)),
\end{equation}
where the action is defined as
\begin{equation}
    \mathcal{S}^{\dagger}_{\textrm{KM}}=\int_{0}^t \diff{\tau}\,\mathrm{i}\theta(\tau)\partial_{\tau} \eta(\tau)-\mathcal{H}^{\dagger}_{\textrm{KM}}(\eta(\tau),\mathrm{i}\theta(\tau))
\end{equation}
As discussed in the main text, we know that the deterministic limit is recovered for $\theta\to0$. To obtain the Langevin expression, we expand $\mathcal{S}^{\dagger}$ around $\mathrm{i}\theta=0$ to second order and integrate with respect $\theta(\tau)$, which leads to
\begin{subequations}
\begin{align}\label{eq:expressP}
    p(n,t) &= \int_{[0}^{t)}\mathcal{D}[\eta] \, 
    e^{-\mathcal{S}^\prime}
    \delta(n-\eta(t)),
    \intertext{where we have defined}
    \mathcal{S}^\prime &= \displaystyle\int_{0}^td\tau\,
    \frac{\left(\partial_{\tau}\eta-\partial_{\mathrm{i}\theta}\mathcal{H}^{\dagger}_{KM}|_{\theta=0}\right)^2}{2\partial^2_{\mathrm{i}\theta}\mathcal{H}_{KM}^{\dagger}|_{\theta=0}},
\end{align}
\end{subequations}
and $\mathcal{D}[\eta]$ is the renormalized measure. 
We recognize from~\cref{eq:expressP} the Feynman-Kac formula~\cite{weber2017master}
\begin{equation}
     p(n,t)=\langle\delta(n-\eta(t)) \rangle_{\eta},
\end{equation}
where $\langle\cdot\rangle_{\eta}$ is the average over the trajectories defined by the following It\^o process
\begin{equation}
     \ddiff{\eta(t)}=\partial_{\mathrm{i}\theta}\mathcal{H}^{\dagger}_{\textrm{KM}}\big|_{\theta=0} + \sqrt{\partial^2_{\mathrm{i}\theta}\mathcal{H}^{\dagger}_{\textrm{KM}}\big|_{\theta=0}}\cdot \xi(t),
\end{equation}
with $\xi(t)$ denoting the standard white noise. We can replace $\mathrm{i}\theta$ with $\theta$ and, including the relation of~\cref{eq:re}, we recover~\cref{eq:efflangevin}. 

\section{The role of initial conditions on phase transitions}\label{sec:app:initialconditions}%
Generally, chemical reaction networks can exhibit highly complex phase portraits with many distinct attractors, especially when the number of compartments grows.
In some cases, this may lead to behavior that affect the procedure described in this work.
Notably, as we are considering systems that are typically out of equilibrium, the systems' stationary state may depend on the initial conditions.
Subsequently, the topology of the phase space may vary in such a way that the asymptotic behavior of the system changes as well.
This, in turn, thus also determines whether a specific phase transition is observed or not, or can change the characterization of the observed phase transition.

\begin{figure}[t]
  \centering
  \includegraphics[width=.7\columnwidth]{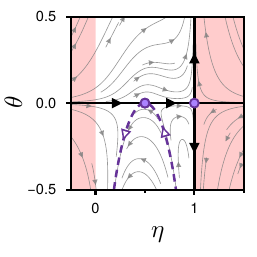}
  \caption{%
    \textbf{Phase plot of a simple $SIS$-like model illustrates dependence on initial conditions.}
    Phase portrait in the $(\eta,\theta)$ plane of the model of~\cref{eq:app:threebodySIS} at the critical point $\alpha = 4\rho$ (see the text).
    Note that when the initial condition $n_S^0 > \tfrac{1}{2}$ (here, equivalent to $\eta=\tfrac{1}{2}$ at $t=0$) the only attractor is $(1,0)$ and no phase transition can be observed.
    Relevant parameters are $N=1$, $\rho=1$ and $\alpha=4$.
  }
  \label{fig:app:SISinitialcond}
\end{figure}

To illustrate the effect that initial conditions can have on a phase transition, let us consider again the modified $SIS$ model (\cref{subsec:modifiedSIS}).
We choose $\gamma=0$ in order to simplify the model, which leaves only the three-body interaction to be the sole mechanism.
That is, our system can be captured by the following elementary reactions
\begin{subequations}
  \begin{align}
    S + 2 I &\xrightarrow{\alpha/N^2} 3I \\
    I &\xrightarrow{\rho} S
  \end{align}
  \label{eq:app:threebodySIS}
\end{subequations}
For this system, one can identify three fixed points in the $(\eta,\theta)$ plane along the mean-field line with $\theta=0$, which for $N=1$ read
\begin{equation}
  F_1 = (1,0) \quad \textrm{and} \quad
  F_{\pm} = \bigg(
            \frac{1}{2} \pm \sqrt{\frac{\alpha - 4\rho}{4\alpha}},
            0
            \bigg)
\end{equation}
The critical condition is, thus, obtained when $F_+ = F_-$, which occurs for $\alpha=4\rho$.
At this point, new fixed points emerge, and we, thus, expect to see a first-order transition.
However, when assuming that $n_S \approx 1$, which indicates the imminent outbreak of a new disease, the phase transition is not observed when numerically integrating the system dynamics.
This can additionally be observed by inspecting the fields in the phase plot (\cref{fig:app:SISinitialcond}).
The lack of phase transition can be understood by realizing that when we vary the rates, we actually determine whether a fixed point attracts or repels.
More specifically, for $\alpha < 4\rho$, we find that $F_1$ is the sole attractor, while for $\alpha > 4\rho$ the fixed point depends on the initial condition.
When $n_S^0 \equiv n_S(t=0) > \tfrac{1}{2}$, $F_1$ remains the sole attractor and, thus, one can conclude that there is no phase transition.
This can be more closely examined by solving for the critical rate $\alpha^\prime$ for which $F_+=(n_S^0,0)$, which reads
\begin{equation}
  \alpha^\prime = \frac{\rho}{n_S^0 ( 1 - n_S^0)}  
\end{equation}
Clearly, when $n_S^0 \approx 1$, we have that $\alpha^\prime \rightarrow \infty$, and, hence, the system effectively does not display a phase transition.
If instead $n_S^0 < \tfrac{1}{2}$ (see~\cref{fig:app:SISinitialcond}), $F_-$ becomes the attractor, and we find a phase transition at $\alpha = 4\rho$.

The simple model discussed here illustrates that initial conditions can play a crucial role in the characterization of phase transitions.
However, it is also clear that our dimension reduction scheme provides a straightforward graphical interpretation of the problem.
Using the phase plot that is obtained by this procedure allows us to intuitively assess under what conditions the initial condition modify the fixed points and, subsequently, the phase transitions of a system.
This stands in stark contrast with other approaches that often need to rely on brute-force inspection of the fixed points, such as numerically integrating the system for a set of initial conditions.

%%%
\section{Identifying phase transitions in a model of tax evaders}\label{sub:tax_app}%
Although the epidemiological $SIS$ model (Section~\ref{subsec:SIS}), and its modification (Section~\ref{subsec:modifiedSIS}) are interesting from an illustrative perspective, they do not warrant a dimension reduction as there is only a single effective compartment, as $N = n_S + n_I$. Here, we treat a more challenging, multicomparment CRN that displays both continuous and discontinuous phase transitions. In particular, we analyze a model for tax evasion introduced elsewhere~\cite{tax}. It includes three compartments, namely, honest individuals $H$, susceptible ones $S$, and evaders $E$. 
The dynamics are described by the following elementary reactions:
\begin{subequations}
    \begin{align}
        E+H&\xrightarrow{\lambda/N} E+S\\
        S&\xrightarrow{\alpha}E\\
        E&\xrightarrow{\beta}H \\
        H+E&\xrightarrow{\delta/N}2H\\
        H+S&\xrightarrow{\epsilon/N} 2H
    \end{align}
\end{subequations}
The DPH is given by
\begin{align}
    \mathcal{H}=&\frac{\epsilon}{N}a_{H}^{\dagger}(a_{H}^{\dagger}-a_{S}^{\dagger})a_{H}a_S+\frac{\delta}{N} a_{H}^{\dagger}(a_{H}^{\dagger}-a_{E}^{\dagger})a_{H}a_E \nonumber \\
    &+ \frac{\lambda}{N}a_{E}^{\dagger}(a_{S}^{\dagger}-a_{H}^{\dagger})a_{H}a_E+\alpha (a_{E}^{\dagger}-a_{S}^{\dagger})a_S \nonumber \\ 
    &+\beta(a_{H}^{\dagger}-a_{E}^{\dagger})a_E,
\end{align}
and the mean-field equations read
\begin{subequations}
    \begin{align}
    \partial_t\langle n_H \rangle &=
    \langle n_H \rangle \left( 
        \frac{\epsilon}{N} \langle n_S \rangle + 
        \frac{\delta}{N} \langle n_E \rangle - 
        \frac{\lambda}{N} \langle n_E \rangle 
    \right)
    + \beta \langle n_E \rangle \\
    \partial_t\langle n_S \rangle &=
    -\langle n_S \rangle \left( \
    \alpha + \frac{\epsilon}{N}\langle n_H \rangle \right) +
    \frac{\lambda}{N} \langle n_H \rangle \langle n_E \rangle \\
    \partial_t\langle n_E \rangle &= 
    -\langle n_E \rangle \left( 
    \beta + \frac{\delta}{N} \langle n_H \rangle
    \right) + \alpha \langle n_S \rangle
 %\partial_t\langle n_H\rangle&=\frac{\epsilon}{N}\langle n_H\rangle\langle n_S\rangle+\beta \langle n_E\rangle+\frac{\delta}{N}\langle n_H\rangle\langle n_E\rangle - \frac{\lambda}{N} \langle n_H\rangle\langle n_E\rangle \\
 %\partial_t\langle n_S\rangle & = -\alpha \langle n_S\rangle-\frac{\epsilon}{N}\langle n_H\rangle\langle n_S\rangle+\frac{\lambda}{N} \langle n_H\rangle\langle n_E\rangle\\
 % \partial_t\langle n_E\rangle&=\alpha \langle n_S\rangle-\beta \langle n_E\rangle-\frac{\delta}{N} \langle n_E\rangle\langle n_H\rangle.
\end{align}
\end{subequations}

\begin{figure*}[t]
    \centering
    \begin{tikzpicture}
        \node[inner sep=0pt,anchor=west] at (0,0) (phase) {%
            \includegraphics[width=.45\linewidth]{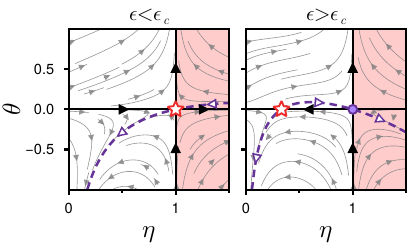}%
        };
        \node[inner sep=0pt, anchor=west,xshift=1cm,yshift=-1.5ex] (phi) at (phase.east) {%
            \includegraphics[width=.4\linewidth]{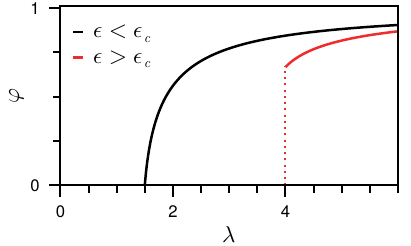}
        };
        \node[anchor=center] (a) at (phase.north west) {\textbf{(a)}};\node[anchor=center] at (a -| phi.north west) {\textbf{(b)}};        
    \end{tikzpicture}
    \caption{%
    \textbf{Second- and first-order phase transition in a model of tax evaders.} 
    The order of the transition depends on the value of $\epsilon$, where for $\epsilon<\epsilon_c$ a second-order transition is observed, and for $\epsilon>\epsilon_c$ a first order.
    Note that the second-order transition was overlooked by the original authors of~Ref.~\citep{tax}.
    \textbf{(a)}~The phase portrait in the Cole-Hopf variables. 
    For more clarification on the phase plot, see~\cref{fig:intro}.
    \textbf{(b)}~The order parameter $\varphi = 1 - \langle n_H \rangle/N$ versus the control parameter $\lambda$.
    Other relevant variables are $\alpha=1$, $\beta=0.5$ and $\delta=0.5$.
    For these variables, we have $\epsilon_c=1$ and $\lambda_1 = 1.5$ for $\epsilon<\epsilon_c$ and $\lambda_2 = 4$ for $\epsilon > \epsilon_c$.
    }%
    \label{fig:taxevasion}
\end{figure*}

In this model, $N$ is fixed and we consider the order parameter $\varphi=1-\langle n_H\rangle/N$. 
After rescaling ($N=1$), we take as first integrals the last mean-field equation and the population conservation, obtaining the parametrization
\begin{equation}
    q_S=\frac{(1-q_{H} x_{H}) (\beta +\delta  q_{H})}{\alpha +\beta +\delta  q_{H}}, \quad 
    q_E=\frac{\alpha -\alpha  q_{H} x_{H}}{\alpha +\beta +\delta  q_{H}}
\end{equation}
The corresponding reduced Hamiltonian subsequently reads
\begin{equation}
    \mathcal{H}^{\star} = \frac{(\eta -1) (1 - e^\theta)}{e^{\theta } (\alpha +\beta )+\delta  \eta } 
    \Big(
    e^{\theta } \left[
    \alpha (\beta + \delta\eta) + \beta\epsilon\eta 
    \right] +
    \eta \left[\delta\eta\epsilon - \alpha\lambda\right]
    \Big),
\end{equation}
and the trivial zero-energy lines are $\eta=1$ and $\theta=0$. The nontrivial zero-energy lines are described by
\begin{equation}
    \theta=\log \left(\frac{\alpha \lambda \eta-\delta \epsilon \eta^2}
    {\alpha\beta + \alpha\delta\eta + \beta\eta\epsilon}\right)
\end{equation}

We denote by $F_{-}$ and $F_{+}$ the intersection of the nontrivial energy curve with the mean-field line, where $\eta_{F_{-}}\le\eta_{F_+}$, and $F_{1}=(1,0)$ is the usual absorbing state (\cref{fig:taxevasion}). 
By considering $\lambda$ as control parameter, we find that $\theta=0$ when
\begin{equation}\label{eq:taxescrit}
    \lambda_c = \frac{(\alpha+\epsilon)(\beta+\delta)}{\alpha}
\end{equation}
For $\lambda<\lambda_c$ the system converges to $F_1$ and there is no phase transition. 
Moreover, at the critical point, which occurs at 
\begin{equation}
    \epsilon_c = \frac{\alpha\beta}{\delta},
\end{equation}
the nontrivial curve is tangent to the mean-field line and $F_{+}=F_{-}=F_1$, yielding a tricritical point. 
For $\epsilon<\epsilon_c$, we find that, at the critical point, $F_{-}=F_1$. 
By studying the eigenvalues, we observe that $F_{-}$ becomes the attractor (\cref{fig:taxevasion}a) and a second-order phase transition occurs. 
Instead, if $\epsilon > epsilon_c$ we see that $F_{+}=F_1$ and the dynamics jumps to the new attractor $F_{-}$.
Therefore, there is a discontinuity of the asymptotic behavior, which indicates a first-order transition (\cref{fig:taxevasion}b).

We point out that, while the critical value $\lambda_c$ corresponds to the result predicted in~\cite{tax}, our method finds a continuous critical transition for low $\epsilon$ that had not been reported before.
This clearly underscores the advantage of our proposal, as it provides a transparent method for detecting critical points that could otherwise be overlooked due to the lack of exploration in certain parameter regions.

%%%%
\section{Details on global attractors and parametrizations of a Lotka-Volterra model}\label{sec:app:spatiallotkavolterra}
As discussed in~\cref{sec:spatiallotkavolterra}, one can overcome parametrization degeneracy through a careful inspection of the dynamics of the system within a specific region of the parameter space.
In practice, this means that one needs to make sure to choose the parametrization that correctly identifies all fixed points, even when some of these fixed points cannot be physically attained.
In this specific model, unphysical fixed points are those that describe negative predator abundances, and, hence, prey abundances that exceed the total number of available patches $N-D$.
While in some cases these attractors are easy to identify, in other cases the dynamics may need to be solved.
Yet, it is important to note that our proposed method alleviates the need to resort to brute-force methods such as numerical integration, as one can identify unphysical fixed points.

\begin{figure*}[t!]
    \centering 
    \includegraphics[width=.975\linewidth,trim={0 1ex 0 0},clip]{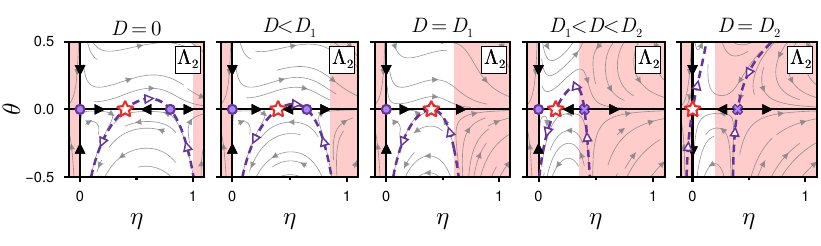}
    \caption{%
        \textbf{Phase portraits of the Lotka-Volterra model after resolving the parametrization degeneracy.}
        Phase portraits in the $(\eta,\theta)$ plane of the Lotka-Volterra model (\cref{eq:spatiallotkavolterra}, see also~\cref{sec:spatiallotkavolterra} for more details) for different amounts of habitat loss.
        Here, we show only the parametrization that resolves the degeneracy, as $\Lambda_2$ is the only parametrization that correctly identifies all fixed points of the system.
        For more clarification on the phase portrait, see~\cref{fig:intro}, and for more details on the model and its parameters, see~\cref{fig:spatiallotkavolterraphaseplot}.
    }%
    \label{fig:app:spatiallotkavolterraparam2}
\end{figure*}

\begin{figure}[t!]
    \centering
    \includegraphics[width=0.95\columnwidth]{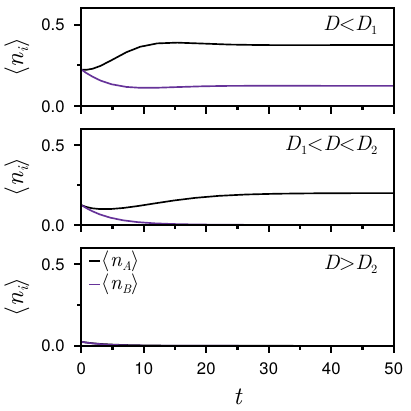}
    \caption{%
        \textbf{Mean-field dynamics of the predator-prey model of~\cref{eq:spatiallotkavolterra}}.
        Curves are shown for the three distinct regions of interest.
        In the first region, $D<D_1$, both prey $A$ and predator $B$ coexist, while in the second region, $D_1 < D < D_2$, the predator goes extinct while the prey species remains.
        In the final third region, $D > D_2$, there are not enough patches to sustain either prey or predator, and both go extinct.
        Note that the steady states obtained by investigating the dynamics are always physical.
        Dynamics are shown for values of the order parameter $D = 0.1$ (top), $D=0.5$ (middle), and $D=0.9$ (bottom).
        All parameters are as in~\cref{fig:spatiallotkavolterraphaseplot}.
    }%
    \label{fig:app:spatiallotkavolterratrajectories}
\end{figure}

As discussed in the main text, we obtain two parametrizations for the model, namely, $\Lambda_1$ and $\Lambda_2$.
Only $\Lambda_2$ correctly identifies all fixed points (\cref{fig:app:spatiallotkavolterraparam2}), but some of them are not within the physical domain.
Close inspection of the mean-field dynamics can additionally be used to garner intuition (\cref{fig:app:spatiallotkavolterratrajectories}).
Indeed, we see that there is only a single global attractor for all values of $D$, and that the nonphysical fixed points for which $\eta > N - D$ cannot be obtained.
Therefore, the (mean-field) dynamics cannot be easily gathered from phase plots that are defined by $\Lambda_2$, yet they can still be used to identify the phase transitions.

Recall that in this model, two phase transitions occur for distinct values of the control parameter $D$ (see~\cref{sec:spatiallotkavolterra} and~\cref{fig:app:spatiallotkavolterraparam2,fig:app:spatiallotkavolterratrajectories}).
For $D=D_1$, we note that the fixed point varies smoothly and that no additional fixed point appears, and, thus, the transition from coexistence to only the prey remaining is of second order.
The same reasoning applies to the critical point at $D=D_2$, and the transition from positive prey abundances to full extinction is also of second order.

Finally note that this specific model, only a single parametrization, and subsequently a single reduced Hamiltonian, remains valid, and, therefore, the degeneracy is completely resolved.

%%%
\section{On the accuracy of the dimension reduction to capture stationary fluctuations}\label{sec:app:reduced}%
As explained in~\cref{sec:reduced}, the dimension reduction process yields reduced Langevin and Fokker-Planck equations. 
The reduced description consistently captures the correct mean-field dynamics of the original system and its critical behavior, if any. 
Nevertheless, the accurate characterization of stochastic behavior is CRN dependent.
In brief, when the CRN includes spontaneous reactions, i.e., there are reactions of the type $A \rightarrow B$, the reduced description may fail to accurately capture equilibrium fluctuations and the transient to stationarity. 
In contrast, when such reactions are absent, the reduced description reliably encapsulates these aspects of the dynamics.
To illustrate this, we examine two distinct models: one derived from epidemiology, which includes spontaneous reactions, and another from ecology, which is free of such reactions.

\subsection{Dynamics with spontaneous reactions: higher-order \texorpdfstring{$SIRS$}{SIS} model}\label{sec:app:reducedoad}
Here, we discuss an epidemiological model that incorporates more realistic features than those studied in ~\cref{subsec:SIS} and \cref{subsec:modifiedSIS}. We use it to illustrate the mismatches that spontaneous reactions introduce.

We consider a modified $SIRS$ model that, compared to a standard $SIRS$, includes an extra spreading mechanism that can be represented by a three-body interaction.
A slight modification of this model has recently been put forward to replicate nonlocal spreading behavior of \emph{cascading failures} in network infrastructures~\cite{scagliarini2025assessing}:
For example, in real-world power grids, failures are known to spread nonlocally~\cite{demarco2001phase,zhao2016spatiotemporal}.
The elementary reactions read
\begin{subequations}\label{eq:app:oad}
  \begin{align}
    S + I &\xrightarrow{\gamma/N} 2I \\
    S + 2I &\xrightarrow{\alpha/N^2} 3I \\
    I &\xrightarrow{\rho} R \\
    R &\xrightarrow{\sigma} S
  \end{align}
\end{subequations}
We proceed as usual by finding the Doi-Peliti Hamiltonian.
The components to freeze are the infected and the recovered ones, and this is equal to specifying the constraints
\begin{subequations}\label{eq:OADconstraints}
  \begin{align}
    x_I = x_R = 1, \\
    \intertext{and}
    q_I = \frac{\sigma(1 - x_S q_S)}{\rho+\sigma}, \qquad
    q_R &= \frac{\rho(1 - x_S q_S)}{\rho+\sigma},
  \end{align}  
\end{subequations}
where the constraints on $q_{I,R}$ have been obtained by investigating the mean-field equations and applying our dimension reduction scheme by setting $\partial_t \langle n_{R} \rangle = 0$.
After applying the Cole-Hopf transform and normalizing by setting $N=1$, we find that the reduced Hamiltonian has trivial zero-energy lines at $\theta = 0$ and $\eta=1$, and that the nontrivial zero-energy line is given by the curve
\begin{equation}
  \theta = \log \left(
    \frac{\eta \left[ \alpha\sigma(1 - \eta) + \gamma(\alpha+\sigma) \right]}{\rho(\rho+\sigma)}
  \right)
\end{equation}
The inspection of the zero-energy lines and their intersections reveals that these display the same structure as the modified $SIS$ model, as expected.
We, therefore, recover the same topology that we reported in~\cref{subsec:modifiedSIS}, and the phase portraits are equivalent.
Briefly, it means that for some $\alpha < \alpha_c$ a second-order transition is observed, while for $\alpha>\alpha_c$ the transition is of first order, where $\alpha_c = \rho(\rho+\sigma)/\sigma$.
Finally, at $\alpha=\alpha_c$, we find a tricritical point.

%% FIGURE
\begin{figure}[t]
  \centering
  \begin{tikzpicture}
      \node[anchor=west,inner sep=0pt] (traj) at (0,0) {%
        \includegraphics[width=.95\linewidth]{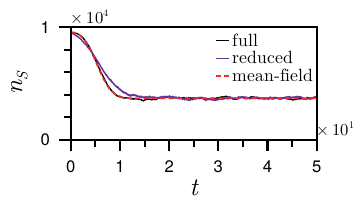}
      };
      \node[anchor=north,inner sep=0pt,yshift=1em] (hist) at (traj.south) {%
        \includegraphics[width=.95\linewidth]{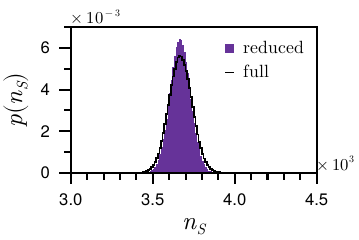}
      };
      \node[anchor=north,inner sep=0pt] (a) at (traj.north west) {(a)};
      \node[anchor=north,inner sep=0pt] (b) at (hist.north west) {(b)};
  \end{tikzpicture}  
  \caption{%
    \textbf{Reduced dynamics and distributions of fluctuations in the higher-order $SIRS$ model}
    (a)~Comparison between typical realizations of the full dynamics, the reduced dynamics, and the mean-field dynamics of the higher-order $SIRS$ model in~\cref{eq:app:oad}.
    While the full dynamics essentially describes fluctuations around the mean-field dynamics, the reduced dynamics typically does not, and follows a slight different evolution in time.    
    (b)~Histograms of the fluctuations of the susceptible compartment.
    Both histograms have been obtained by numerical integration of the stochastic dynamics using an Euler-Maruyama scheme with $\Delta t = 0.01$, and have been computed after discarding the first $10^3$ steps.
    Note that only the first moments of the distributions coincide, indicating that the reduced dynamics does not go beyond the mean field for this specific model (see the text).
    The plots are created with $N=10^4$, $\gamma=1$, $\alpha=0.2$, $\rho=0.4$ and $\sigma=1$.
  }
  \label{fig:app:sirshist}
\end{figure}

Because the original system includes spontaneous reactions, the resulting correlation matrix of the Langevin equation contains off-diagonal elements, leading to some expected discrepancies between the full multicompartmental description and the reduced one. 
Although the expression for the stochastic dynamics of $\eta$ is lengthy, the drift and noise term can be written as functions of the system parameters, and can be numerically integrated by using, for example, an Euler-Maruyama scheme. 
Upon performing such an integration (see Fig.~\ref{fig:app:sirshist}), we confirm that the time evolution of the reduced dynamics deviates from that of the full system.
Notably, in this specific case, the reduced description yields a transient curve with a decay timescale of the same order of magnitude as that of the full system, and the largest mismatch in the histogram of $n_S$ is approximately 10\% at its peak. 
While these results may appear acceptable, we anticipate that the level of mismatch depends on the specific CRN and is likely to increase with the number of components. 
A comprehensive characterization of these trends lies beyond the scope of this work, but it merits further investigation in future studies.

\subsection{Dynamics without spontaneous reactions: Generalized Lotka-Volterra model}\label{sec:app:reducedglv}%
Let us now illustrate
the opposite scenario, namely, a CRN with a large number of compartments but without spontaneous reactions. As expected from our theoretical derivations, the reduced description should perfectly approximate the one of the full system.

Let us take a generalized Lotka-Volterra model with $M$ distinct species.
The (chemical) Langevin equation of the generalized Lotka-Volterra model is known, and reads~\cite{giuli2022dynamical}
\begin{equation}
  \dv{n_i}{t} = \frac{r_i n_i}{\kappa_i} \left( \kappa_i - n_i - \sum_{j\neq i} \alpha_{ij} n_j \right)
  + \sqrt{n_i} \cdot \xi_i, 
\end{equation}
where $r_i$ and $\kappa_i$ are, respectively, the growth rate and the carrying capacity of species $i$, and $\alpha_{ij}$ are the interaction coefficients.
Note that the noise function depends only on $n_i$ itself. This is the key fact that renders the reduced dynamics an incredibly good approximation for the full stochastic system.

When writing down the elementary reactions, one should be mindful of the fact that the deterministic part of this model contains a quadratic term in $n_i$.
Following the steps highlighted in Ref.~\cite{giuli2022dynamical}, the reactions are
\begin{subequations}
  \begin{align}
    N_i &\xrightarrow{r_i} 2N_i \\
    N_i + N_i &\xrightarrow{r_i/\kappa_i} N_i \\
    N_i + N_j &\xrightarrow{\alpha_{ij}/\kappa_i} N_j
  \end{align}
\end{subequations}
and, thus, the Doi-Peliti Hamiltonian of such a system is
\begin{align}
  \label{eq:glvdph}
  \begin{split}
    \mathcal{H} &= \sum_i r_i a_i^\dagger (a_i^\dagger - 1) a_i +
                  \sum_i \frac{r_i}{\kappa_i} a_i^\dagger (1 - a_i^\dagger ) a_i a_i
                   \\ 
                &\quad + \sum_{i,j} \frac{\alpha_{ij}}{\kappa_i} a_j^\dagger (1 - a_i^\dagger) a_i a_j
  \end{split}
\end{align}
The final sum can be reconciled with the sum within the stochastic differential equation (which excludes $j=i$) simply by considering $\alpha_{ii} = 0$, or by appropriate scaling of diagonal elements of the interaction matrix.
Here, we choose the former.

By again using the field transformations $a_i^\dagger \Leftrightarrow x_i$ and $a_i \Leftrightarrow q_i$, we can readily apply our dimension reduction formalism and, without loss of generality, freeze all compartments but $i=1$.
Then, the reduced Hamiltonian $\mathcal{H}^\star$ is the Hamiltonian evaluated at $x_j = 1$ and $q_j = Q(q_1,\boldsymbol{x})$, where $Q$ follows from the constraints.
The constraints to determine $q_j$ are obtained from the mean-field equations for $\langle n_j \rangle$, i.e.,
\begin{equation}
  \partial_t \langle n_j \rangle =
  \left[\frac{\partial \mathcal{H}(\boldsymbol{x}, \boldsymbol{q})}{\partial x_j}\right]_{\boldsymbol{x}=1}
\end{equation}
Recalling that we chose to freeze all compartments but the first, we solve the mean-field equations for $q_j$ after setting $\partial_t \langle n_j \rangle = 0$.
Generally, the parametrizations of the form $q_j = Q(q_1,\boldsymbol{x})$ are lengthy but can be solved either numerically or symbolically.

\begin{figure}[t!]
  \centering
  \begin{tikzpicture}
      \node[anchor=west,inner sep=0pt] (traj) at (0,0) {%
        \includegraphics[width=.95\linewidth]{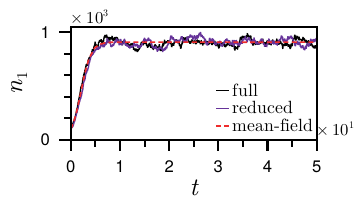}
      };
      \node[anchor=north,inner sep=0pt,yshift=1em] (hist) at (traj.south) {%
        \includegraphics[width=.95\linewidth]{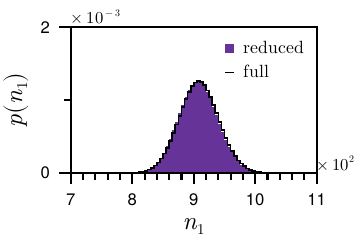}
      };
      \node[anchor=north,inner sep=0pt] (a) at (traj.north west) {(a)};
      \node[anchor=north,inner sep=0pt] (b) at (hist.north west) {(b)};
  \end{tikzpicture}
  \caption{%
    \textbf{Reduced dynamics and distributions of fluctuations in the generalized Lotka-Volterra model}
    (a)~Comparison between typical realizations of the full dynamics, the reduced dynamics, and the mean-field dynamics of the generalized Lotka-Volterra model with $M=24$ species.
    In this case, both the full (stochastic) dynamics and the reduced dynamics are essentially fluctuations around the mean-field dynamics.
    In addition, both the first and second moments of distribution of fluctuations capture the full dynamics.
    (b)~Histograms of the fluctuations of the first species $n_1$.
    Interactions are such that the mean-field describes a feasible equilibrium (see text).
    Note that the distribution of fluctuations of the reduced dynamics is indistinguishable from that of the full dynamics.
    Both histograms have been obtained by numerical integration of the stochastic dynamics using an Euler-Maruyama method with $\Delta t = 0.01$.
    We have integrated the system up to $t=5\cdot 10^5$, and the histograms have been computed after discarding the first $10^4$ steps.
    Positive interaction coefficients were randomly sampled with the same seed for both the full system and the reduced one.
    We set $r=1$ and $\kappa=10^3$ for all species, and $\mu_A=1$ and $\sigma_A=0.1$ for the interaction coefficients.
    On average, 10\% of the species interacted --- i.e., $c=0.1$ --- hence $\boldsymbol{\alpha}$ could be considered sparse, and we were able to compute the drift and noise term for the reduced dynamics symbolically.
  }%
  \label{fig:app:glvhist}
\end{figure}

In the case of the generalized Lotka-Volterra, one needs to take some additional care by ensuring that the parameters define a \emph{feasible} system, i.e., a system where each $n_i>0$ up until large $t$.
If this is not the case, then some $q_j$ must equal zero, yet one would need to inspect the full (mean-field) dynamics in order to know which species go extinct.
Effectively, this means that the approach described above holds in the weak-interaction limit~\cite{bunin2017ecological,biroli2018marginally,altieri2021properties}.
More specifically, one considers interactions to occur with some probability $c$, and some strength that is sampled from a distribution with finite moments.
Typically, a Gaussian distribution is chosen for the interaction coefficients, 
\begin{equation}
  \alpha_{ij} = \frac{\mu_A}{M} + \frac{\sigma_A z_{ij}}{\sqrt{M}},
\end{equation}
where $z_{ij}$ is a standard normal variable with mean $0$ and variance $1$.
While a thorough investigation of the generalized Lotka-Volterra model is well beyond the scope of this work (but see, e.g., Refs.~\cite{akjouj2023equilibria,galla2024generatingfunctional} for pedagogical reviews on the topic), the essence is that our dimension reduction scheme applies to scenarios in which $\mu_A$ and $\sigma_A$ define systems with a unique feasible equilibrium that is globally attracting~\cite{bunin2017ecological, biroli2018marginally}.

Putting these technicalities aside, we proceed with the pipeline of our dimension reduction. The reduced dynamics are obtained by applying the Cole-Hopf transformation on the reduced Hamiltonian and deriving the drift and noise functions, the reduced Langevin equation is
\begin{equation}
  \label{eq:langevinlv}
  \dv{\eta}{t} = \partial_\theta \mathcal{H}^\star_0 +\sqrt{\partial^2_\theta \mathcal{H}^\star_0} \cdot \xi_t, 
\end{equation}
where $\partial^k_\theta \mathcal{H}^\star_0 = [\partial^k_\theta \mathcal{H}^\star]_{\theta=0}$. Because of the absence of spontaneous reactions, the diffusion term is diagonal, and we expect that Eq.~\eqref{eq:langevinlv} yields an accurate approximation of the full dynamics.

Indeed, when observing the histograms of the equilibrium fluctuations, we show in~\cref{fig:app:glvhist} that both the time evolution and the fluctuation distributions, of the full and the reduced dynamics are essentially indistinguishable, even if we have chosen a large number of compartments $M = 24$.
In fact, upon close inspection, the first two moments are exactly equal up to numerical precision, a clear indication that the reduced dynamics captured phenomena well beyond the mean-field dynamics.

\end{document}